\documentclass[iop]{emulateapj}

\newcommand{\midline}{\, | \,}
\newcommand{\betaeff}{\beta_\mathrm{eff}}
\newcommand{\dusttemp}{T_\mathrm{eff,d}}
\newcommand{\kappaeff}{\kappa_{\mathrm{eff},160}}

\slugcomment{Accepted for publication in the ApJ.}
\begin{document}

\shortauthors{Gordon et al.}
\shorttitle{HERITAGE LMC/SMC Dust}

\title{Dust and Gas in the Magellanic Clouds from the HERITAGE
  {\it Herschel} Key Project. \\
  I. Dust Properties and Insights into the Origin of the Submm Excess
  Emission\footnote{{\it Herschel} is an ESA space observatory with science instruments provided
by European-led Principal Investigator consortia and with important participation from NASA.}}

\author{Karl~D.~Gordon\altaffilmark{1,2}, 
   Julia~Roman-Duval\altaffilmark{1},
   Caroline~Bot\altaffilmark{3},
   Margaret~Meixner\altaffilmark{1},
   Brian~Babler\altaffilmark{4},
   Jean-Philippe~Bernard\altaffilmark{5,6},
   Alberto~Bolatto\altaffilmark{7},
   Martha~L.~Boyer\altaffilmark{8,9},
   Geoffrey~C.~Clayton\altaffilmark{10},
   Charles~Engelbracht\altaffilmark{11, 12},
   Yasuo~Fukui\altaffilmark{13},
   Maud~Galametz\altaffilmark{14},
   Frederic~Galliano\altaffilmark{15},
   Sacha~Hony\altaffilmark{15},
   Annie~Hughes\altaffilmark{16},
   Remy~Indebetouw\altaffilmark{17},
   Frank~P.~Israel\altaffilmark{18},
   Katie~Jameson\altaffilmark{7},
   Akiko~Kawamura\altaffilmark{19},
   Vianney~Lebouteiller\altaffilmark{15},
   Aigen~Li\altaffilmark{20},
   Suzanne~C.~Madden\altaffilmark{15}, 
   Mikako~Matsuura\altaffilmark{21},
   Karl~Misselt\altaffilmark{11},
   Edward~Montiel\altaffilmark{10,11},
   K.~Okumura\altaffilmark{15},
   Toshikazu~Onishi\altaffilmark{22},
   Pasquale~Panuzzo\altaffilmark{15,23},
   Deborah~Paradis\altaffilmark{5,6},
   Monica~Rubio\altaffilmark{24},
   Karin~Sandstrom\altaffilmark{11},
   Marc~Sauvage\altaffilmark{15},
   Jonathan~Seale\altaffilmark{25, 1},
   Marta~Sewi{\l}o\altaffilmark{25},  
   Kirill~Tchernyshyov\altaffilmark{25}, \&
   Ramin~Skibba\altaffilmark{26, 11}
   }
\altaffiltext{1}{Space Telescope Science Institute, 3700 San Martin
  Drive, Baltimore, MD, 21218, USA}
\altaffiltext{2}{Sterrenkundig Observatorium, Universiteit Gent,
              Gent, Belgium}
\altaffiltext{3}{Observatoire astronomique de Strasbourg, Universit\'e
  de Strasbourg, CNRS, UMR 7550, 11 rue de l~Universit\'e, F-67000
  Strasbourg, France} 
\altaffiltext{4}{Department of Astronomy, 475 North Charter St.,
  University of Wisconsin, Madison, WI 53706, USA}
\altaffiltext{5}{CESR, Universit\'e de Toulouse, UPS, 9 Avenue du
  Colonel Roche, F-31028 Toulouse, Cedex 4, France} 
\altaffiltext{6}{Universit\'e de Toulouse, UPS-OMP, IRAP, 31028
  Toulouse Cedex 4, France} 
\altaffiltext{7}{Department of Astronomy,  Lab for Millimeter-wave
  Astronomy, University of Maryland,  College Park, MD 20742-2421, USA} 
\altaffiltext{8}{Observational Cosmology Lab, Code 665, NASA Goddard
  Space Flight Center, Greenbelt, MD 20771, USA}
\altaffiltext{9}{Oak Ridge Associated Universities (ORAU), Oak Ridge,
  TN 37831, USA}
\altaffiltext{10}{Louisiana State University, Department of Physics \&
  Astronomy, 233-A Nicholson Hall, Tower Dr., Baton Rouge, LA
  70803, USA}
\altaffiltext{11}{Steward Observatory, University of Arizona, 933 North
  Cherry Ave., Tucson, AZ 85721, USA} 
\altaffiltext{12}{Raytheon Company, 1151 East Hermans Road, Tucson, AZ
  85756, USA} 
\altaffiltext{13}{Department of Physics, Nagoya University, Furo-cho,
  Chikusa-ku, Nagoya 464-8602, Japan}
\altaffiltext{14}{European Southern Observatory,
  Karl-Schwarzschild-Str. 2, D-85748 Garching-bei-München, Germany}
\altaffiltext{15}{CEA, Laboratoire AIM, Irfu/SAp, Orme des Merisiers,
  F-91191 Gif-sur-Yvette, France} 
\altaffiltext{16}{Max-Planck-Institut f\"ur Astronomie, K\"onigstuhl 17,
  D-69117 Heidelberg, Germany}
\altaffiltext{17}{Department of Astronomy, University of Virginia,
  and National Radio Astronomy Observatory,  520 Edgemont Road,
  Charlottesville, VA 22903, USA}
\altaffiltext{18}{Sterrewacht Leiden, Leiden University, P.O. Box
  9513, NL-2300 RA Leiden, The Netherlands} 
\altaffiltext{19}{ National Astronomical Observatory of Japan,
  Osawa, Mitaka, Tokyo, 181-8588,  Japan}
\altaffiltext{20}{314 Physics Building, Department of Physics and
  Astronomy, University of Missouri, Columbia, MO 65211, USA} 
\altaffiltext{21}{Department of Physics and Astronomy, University
  College London, Gower Street, London WC1E 6BT, UK}
\altaffiltext{22}{Department of Astrophysics, Graduate School of
  Science, Osaka Prefecture University, Sakai, Osaka 599-8531, Japan}
\altaffiltext{23}{CNRS, Observatoire de Paris - Lab. GEPI, Bat. 11, 5,
  place Jules Janssen, 92195 Meudon CEDEX, France}
\altaffiltext{24}{Departamento de Astronom\'{\i}a, Universidad de
  Chile, Casilla 36-D, Santiago, Chile}
\altaffiltext{25}{The Johns Hopkins University, Department of Physics
  and Astronomy, 366 Bloomberg Center, 3400 N. Charles Street,
  Baltimore, MD 21218, USA}
\altaffiltext{26}{Center for Astrophysics and Space Sciences,
  Department of Physics, University of California, 9500 Gilman Dr, La
  Jolla, San Diego, CA 92093, USA}

\begin{abstract} 
The dust properties in the Large and Small Magellanic Clouds are
studied using the HERITAGE {\it Herschel} Key Project photometric data
in five bands from 100 to 500~\micron.  Three simple models of dust
emission were fit to the observations: a single temperature blackbody
modified by a power-law emissivity (SMBB), a single temperature
blackbody modified by a broken power-law emissivity (BEMBB), and two
blackbodies with different temperatures, both modified by the same
power-law emissivity (TTMBB).  Using these models we investigate the
origin of the submm excess; defined as the submillimeter (submm)
emission above that expected from SMBB models fit to observations $<
200\micron$.  We find that the BEMBB model produces the lowest fit
residuals with pixel-averaged 500~\micron\ submm excesses of 27\% and
43\% for the LMC and SMC, respectively.  Adopting gas masses from
previous works, the gas-to-dust ratios calculated from our the fitting
results shows that the TTMBB fits require significantly more dust than
are available even if all the metals present in the interstellar
medium (ISM) were condensed into dust.  This indicates that the submm
excess is more likely to be due to emissivity variations than a second
population of colder dust.  We derive integrated dust masses of $(7.3
\pm 1.7) \times 10^5$ and $(8.3 \pm 2.1) \times 10^4$~$M_\sun$ for the
LMC and SMC, respectively.  We find significant correlations between
the submm excess and other dust properties; further work is needed to
determine the relative contributions of fitting noise and ISM physics
to the correlations.
\end{abstract}

\keywords{infrared: galaxies, infrared: ISM, ISM: general, Magellanic Clouds}

\section{Introduction}
\label{sec_intro}

Among nearby galaxies, the Large Magellanic Cloud (LMC) and Small
Magellanic Cloud (SMC) represent unique astrophysical laboratories for
interstellar medium (ISM) studies.  Both Clouds are relatively nearby,
the LMC at $\sim$50 kpc \citep{Walker12} and the SMC at $\sim$60 kpc
\citep{Hilditch05}, and provide ISM measurements that are relatively
unconfused along the line-of-sight as compared to similar
observations in the Milky Way (MW).  The LMC and SMC ultraviolet dust
extinction properties show strong variations both internally and in
global averages in comparison to each other and the MW \citep{Lequeux82, Prevot84,
Clayton85, Fitzpatrick85, Gordon03, MaizAppellaniz12}.  The two Clouds
span an important metallicity range with the LMC at $\sim$1/2
Z$_{\sun}$ \citep{Russell92} being above and the SMC at $\sim$1/5
Z$_{\sun}$ \citep{Russell92} being below the threshold of 1/3--1/4
Z$_{\sun}$ where the properties of the ISM change significantly as
traced by the reduction in the Polycyclic Aromatic Hydrocarbon (PAH)
dust mass fractions and (possibly) dust-to-gas ratios \citep{Draine07}.
The far-infrared (FIR) to submillimeter (submm) emission from the
Clouds shows more submm emission than expected from existing dust
grain models, with the SMC having a larger amount of this excess
emission \citep{Israel10, Bot10}.

The submm excess was seen first in the MW using the COBE/FIRAS
\citep{Boggess92, Mather93} observations of high-latitude cirrus dust
emission \citep{Wright91, Reach95}.  These works found the
100--300~\micron\ observations were well modeled with a single
temperature blackbody modified with a power law emissivity, but that
the longer wavelength observations ($\lambda > 300~\micron$) required
a second dust component with a temperature of 4--7~K.  The spatial
correlation of this second dust component with the hotter main dust
component along with physical arguments on dust heating led
\citet{Reach95} to argue that emissivity variations away from a simple
power law 
were more likely to explain the observations than a second component
of very cold dust. 
The need for a non-trivial FIR to submm dust emissivity shape was
quantified by \citet{Li01} where they modified the emissivity of
``astronomical'' silicate grains to have an emissivity with a
shallower wavelength dependence at $\lambda > 200~\micron$ than at
$\lambda < 200~\micron$.  More recently, \citet{Paradis12} analyzed
{\it Herschel Space Observatory} \citep{Pilbratt10} observations of
the MW plane and found a significant submm excess at 500~\micron\ that
increased from the inner to the outer Galaxy.

Previous work on the submm excess in nearby galaxies by
\citet{Galliano03, Galliano05} and \citet{Galametz11} used the combination
of FIR observations ($\lambda < 200~\micron$) from the {\it Infrared
Space Observatory} \citep{Kessler96} and {\it Spitzer Space Telescope}
\citep{Werner04Spitzer} with submm observations ($\lambda \sim
850~\micron$) taken using ground-based observatories.  These works provided
strong evidence of a submm excess at $\sim$850~\micron\ and that this
excess is largest in low metallicity galaxies.  With the advent of
{\it Herschel} observations, the presence of a submm excess at
500~\micron\ has been established in many low metallicity galaxies
including the Magellanic Clouds \citep{Gordon10, Meixner10,
Galliano11, Dale12, Kirkpatrick13, RemyRuyer13}.

The definition of the submm excess has not been uniformly defined in
the literature, complicating the comparisons between different studies.
Generally, a model is used to define the zero submm excess baseline;
this model varies from simple modified blackbodies to more complex
dust grain models.  In addition, the uncertainties assumed on the
observations have varied leading to the same submm excess level being
considered significant by one work and not significant by another.
This illustrates the need for a uniform definition of reference
spectral energy distribution (SED) from which to measure the submm
excess and a common set of assumptions on the observational
uncertainties.  It is also critically important to properly include
the full observational uncertainties, both correlated and
uncorrelated, as shown by \citet{Galliano11} and \citet{Veneziani13}.  

For clarity in this paper, we adopt the definition of the submm excess
as the excess emission seen at submm wavelengths above that expected
for dust grains with a single temperature and a $\lambda^{-\betaeff}$
emissivity law.  This simple model is used to fit an observed SED,
with the value of $\betaeff$ providing a measure of the effective
emissivity law.  The origin of the observed effective emissivity law
variations may be due to one or a combination of factors including
intrinsic dust emissivity variations, mixing of different dust
compositions, and variations in dust temperatures along the line of
sight.

Laboratory studies of the two main interstellar dust analogs have
shown that carbonaceous grains have $\beta \sim 1-2$
\citep{Mennella95, Zubko96, Jager98} and silicate grains have $\beta
\sim 2$ \citep{Mennella98, Boudet05, Coupeaud11} in the FIR and submm
wavelength range.  The value of $\betaeff$ for a mixed composition
dust population is determined by both the actual ratio of the two
compositions and the spectral shape of the heating radiation field.
Silicate and carbonaceous grains have significantly different
ultraviolet/optical absorption properties and any change in the
radiation field spectrum will change the luminosity weighting present
in the infrared (IR) dust emission SED.  Deviations from simple
$\lambda^{-\beta}$ emissivity laws and dependence on temperature are
seen in laboratory work on dust analogs, with silicate grains having
larger such variations than carbonaceous grains \citep{Mennella98,
Boudet05, Coupeaud11}.  Such deviations have already been seen in
astronomical observations, leading \cite{Li01} to modify their
model of "astronomical" silicates such that it already includes a
submm excess of 11\% at 500~\micron, according to our definition above. 
Similar broken power law dust emissivities have been implied by FIR to
submm observations of the different phases of the MW ISM
\citep{Paradis09}.

Multiple dust temperatures along the line-of-sight can also cause
effective emissivity law variations.  The simplest case to consider is
two dust populations with the second population having a significantly
colder temperature than the first.  Fitting the composite SED of this
dust with a single temperature $\lambda^{-\betaeff}$ emissivity law
model will result in a submm excess at the wavelengths where the
second cold dust population contributes.  Such two temperature models
have been studied by \citet{Juvela12} who find that the $\betaeff$ can
either be higher or lower than the intrinsic $\beta$ depending the
distribution of temperatures.  More complex temperature mixing has
been investigated with similar results \citep{Shetty09a, Shetty09b,
Juvela12, Ysard12}.

The implications for our understanding of dust grain properties are
quite different depending on the origin of the submm excess.  If the
submm excess is due to very cold dust, then the total dust mass would
potentially increase significantly as a large mass of cold dust
is needed to reproduce the observed emission
\citep[e.g.,][]{Galliano05}.  On the other hand, if the submm excess
is due to dependencies of the effective emissivity law with
wavelength, then this provides insights into variations in the ratio
of silicate/carbonaceous grains and/or variations in spectral shape of the
illuminating radiation field.

The Magellanic Clouds provide two of the best laboratories to study
the submm excess given their proximity and lower than MW
metallicities.  Work on this topic in the Magellanic Clouds prior to
the {\it Herschel} observations has used ground-based submm
observations \citep[e.g.,][]{Bot10_Laboca} or low spatial resolution
PLANCK observations.  In particular, the studies by \citet{Israel10} and
\citet{Bot10} clearly show a submm excess in both Clouds, even though
the works were focused on the longer wavelength emission of the Clouds.
They found that the observed submm excess
can be explained using \citet{Draine07model} models with cold
dust grains, but not by emission due to spinning grains, which is the 
likely origin of the excess emission they observed at millimeter to
centimeter wavelengths.  Similar results for the submm 
excess in the SMC were found using the PLANCK observations
\citep[][Verdugo et al.  submitted]{PlankMCs11}.  
In apparent conflict with these wide-field and/or global studies of
dust emission in the Clouds, a spatially resolved study by
\citet{Galametz13} found no evidence for a submm excess at
870~\micron\ in N159, a 
massive star-forming complex in the LMC. As noted by the authors,
however, their conclusions apply only to high surface brightness
regions that can be detected using ground-based submm observations. 

The HERschel Inventory of The Agents of Galaxy Evolution (HERITAGE) in
the Magellanic Clouds {\it Herschel} Key Project has mapped both
Clouds providing observations at 100, 160, 250, 350, and 500~\micron\
\citep{Meixner13}.  The HERITAGE wavelength coverage
(100--500~\micron) and spatial resolution ($\sim$10~pc at 500~\micron)
is well suited to measuring the spatial variations of dust properties
probed by FIR and submm emission.  In particular, these observations
are ideally suited to investigating the nature of the submm excess and
how it varies spatially in each Cloud.  The HERITAGE project test
observations of a strip in the LMC have been analyzed and a measurable
submm excess at 500~\micron\ was found using both simple single
temperature blackbodies \citep{Gordon10} and a more complex dust grain
model \citep{Meixner10, Galliano11}.  These studies found that this
submm excess was anti-correlated with ISM (gas or dust) surface
density.

The goal of this paper is to investigate the submm excess in both
Magellanic Clouds using the full HERITAGE data using simple dust
emission models based on one or two modified blackbodies.  We choose
to use such models for this paper since they allow large potential
variations in the effective emissivity laws, whereas existing dust
grain models do not incorporate the full range of variations indicated
by laboratory studies of ISM dust analogs. 
In addition, we are careful to use a robust model of the
uncertainties in the measurements, including the correlations between
the different {\it Herschel} bands due to the absolute flux
calibration and the background subtraction.  Preliminary versions of
the dust surface density maps derived in this paper were used to study
the correlation between dust and stellar properties in the Magellanic
Clouds by \citet{Skibba12}.  

\section{Data}
\label{sec_data}

The FIR and submm observations of the Magellanic Clouds analyzed in
this study were taken as part of the HERITAGE Key Project
\citep{Meixner13} using the PACS \citep{Poglitsch10} and SPIRE
\citep{Griffin10} instruments on the {\it Herschel Space Observatory}.
The observations provided images of the LMC and SMC at 100, 160, 250,
350, and 500~\micron\ that cover the entire IR emitting regions of
both galaxies ($8^\circ$$\times$$8.5^\circ$ and $5^\circ$$\times$$5^\circ
+ 4^\circ$$\times$$3^\circ$ for the LMC and SMC, respectively).  The
observation and data 
reduction details can be 
found in \citet{Meixner13}.  It is useful to note that as part of the
data reduction, the IRAS 100~\micron\
\citep{Schwering89,Schwering89LMC} and MIPS~160~\micron\ images
\citep{Meixner06, Gordon11} for each galaxy were used to correct for
the drifting baseline of the PACS bolometers.  Thus the PACS 100 and
160~\micron\ images contain the IRAS 100 and MIPS 160 information as
well as the new PACS observations.

Additional processing steps were performed for this study to create
images that had the same spatial resolution and the same
foreground/background subtraction.  First, each image was convolved with a
kernel that transformed the spatial resolution of the images to the
lowest resolution of the set of images, set by the SPIRE 500~\micron\
point-spread-function (PSF) which has a resolution of
$\sim$40\arcsec.  The \citet{Aniano11} convolution kernels were used
for this step as they directly and optimally transform the native PSF
to that of the SPIRE 500~\micron\ PSF.

Second, a foreground subtraction was done to remove the structured emission
due to MW dust (cirrus) emission.  The detailed structure of the MW
dust emission in the PACS and SPIRE bands was predicted using the
integrated MW velocity HI gas maps in the direction of the LMC
\citep{Staveley-Smith03} and SMC \citep{Stanimirovic00, Muller03} and
the \citet{Desert90} model for the local interstellar radiation
field. This model gives the conversion between HI column and infrared
emission.  The conversion coefficients used were 1.073, 1.848, 1.202,
0.620, and 0.252 (MJy/sr) ($1 \times 10^{20}$ H I atoms/cm$^2$)$^{-1}$
for 100, 160, 250, 350, and 500~\micron, respectively.  These
conversion coefficients are higher than those that would be obtained
with the newer DustEM model \citep{Compiegne11} for the same radiation
field, but are similar to the observed correlations between the MW
velocity integrated HI and the diffuse emission measured in the same
bands in regions outside of the SMC.  This step was particularly
important for the SMC where structures with similar surface
brightnesses to those in the galaxy were removed by this subtraction.

Finally, residual large scale structure in the background was removed
using a low order 2D surface polynomial interpolation that was
constrained by regions external to each galaxy.  The baseline
subtraction reduction step for PACS and SPIRE data used different
assumptions for these external regions \citep{Meixner13} and, thus,
this final step ensures that all the images have the same background
subtraction.  This background subtraction is especially important for
the LMC where the SPIRE observations included emission near the edges
of the HERITAGE coverage due to the very extended nature of the LMC
(especially south of the LMC main body) and the excellent sensitivity
of the SPIRE instrument.

\section{Models}
\label{sec_models}

We use three different models to fit the FIR/submm surface
brightness measurements.  The first model is a single temperature blackbody
modified by a single power law emissivity (SMBB).  The second model
assumes the submm excess emission is due to variations in the
wavelength dependence of the dust emissivity law that is parametrized
by a broken power law (BEMBB).  The third model assumes the submm excess
emission is due to a second, lower temperature population of dust
grains (TTMBB).  All our models assume equilibrium heating only and so
we restrict our fits to using only data $\geq$$100~\micron$.  It is
reasonable to expect that the emission at these wavelengths is
dominated by equilibrium emission from dust grains.  In this analysis,
any residual 100~\micron\ contribution due to emission from
transitionally heated grains will yield a somewhat higher dust
temperature (and thus a smaller dust column density) than would be
found with our models. In the great majority of sight lines, this
contribution is too small to be of concern, but may introduce a
systematic bias in the regions near intense star formation.

In general, the surface brightness of dust with temperature, $T_d$, is
\begin{eqnarray}
S_{\lambda} & = & \tau_{\lambda}B_{\lambda}(T_d) \\
 & = & N_d \pi a^2 Q_{\lambda} B_{\lambda}(T_d) \\
 & = & \frac{\Sigma_d}{m_d} \pi a^2 Q_{\lambda} B_{\lambda}(T_d) \\
 & = & \frac{\Sigma_d}{\slantfrac{4}{3}a^3 \rho} \pi a^2 Q_{\lambda} B_{\lambda}(T_d) \\
 & = & \frac{3}{4a\rho} \Sigma_d Q_{\lambda} B_{\lambda}(T_d) \\
 & = & \kappa_\lambda \Sigma_d B_\lambda \label{eq_fulltheory}
\end{eqnarray}
where $\tau_\lambda$ is the dust optical depth, $N_d$ is the dust
column density, $a$ is the grain radius, $Q_\lambda$ is the dust
emissivity, $B_\lambda$ is the Planck function, $\Sigma_d$ is the dust
surface mass density, $m_d$ is the mass of a single dust grain, $\rho$
is the grain density, $\kappa_\lambda$ is the grain
absorption cross section per unit mass.  These equations can be
evaluated in standards units (e.g. cgs or MKS).  We found it
convenient to express $\Sigma_d$ in
$\mathrm{M}_\sun~\mathrm{pc}^{-2}$, $\kappa_\lambda$ in 
$\mathrm{cm}^2~\mathrm{g}^{-1}$, and $B_{\lambda}$ and $S_{\lambda}$
in $\mathrm{MJy} 
\mathrm{sr}^{-1}$ and then Eq.~\ref{eq_fulltheory} is
\begin{equation} 
S_{\lambda} = (2.0891 \times 10^{-4}) \kappa_\lambda \Sigma_d B_\lambda.
\end{equation}

From Eq.~\ref{eq_fulltheory}, it is clear that the values of
$\kappa_\lambda$ and $\Sigma_d$ are completely degenerate.  Without
further information FIR to submm SED observations only constrain
$\tau_\lambda = \kappa_\lambda \Sigma_d$.  Breaking this degeneracy is
possible in the one environment where we have measurements of the
expected amount of dust independent from the measured FIR to submm
dust emission.  This environment is the MW diffuse ISM where
ultraviolet and optical gas-phase absorption measurements provide a
strong constraint on the depletions in the ISM
\citep[e.g.,][]{Jenkins09}.  We use these measurements to calibrate
$\kappa_\lambda$ in \S\ref{sec_kappa_cal} for the models introduced
below.  This calibration ensures that our models produce the right
$\Sigma_d$ in the one place where we know the correct value from
independent measurements.

\subsection{SMBB: Simple Emissivity Law Model}
\label{sec_smbb}

The SMBB predicts the surface brightness assuming a
dust population with single dust temperature modified by 
a simple emissivity law \citep{Hildebrand83}.  The adopted emissivity law is
\begin{equation}
\kappa_\lambda = \frac{\kappaeff^\mathrm{S}}{160^{-\betaeff}} \lambda^{-\betaeff}.
\end{equation}
The value of $\kappaeff^\mathrm{S}$ is set by
fitting of the diffuse MW SED (\S\ref{sec_kappa_cal} and
Table~\ref{tab_mw_diffuse}).  The full set 
of fit parameters for the SMBB model are   
$\theta_\mathrm{S} = (\Sigma_d, \dusttemp, \betaeff)$.   The values
for the dust properties are effective values due to composition and temperature
mixing along the line-of-sight and are not directly comparable to
interstellar dust grain analogs studied in the laboratory (see \S\ref{sec_intro}).

\subsection{BEMBB: Broken Emissivity Law Model}
\label{sec_bembb}

The BEMBB predicts the surface brightness assuming a
dust population with a single dust temperature modified by 
a broken emissivity law.  The adopted emissivity law is
\begin{equation}
\kappa_\lambda = \frac{\kappaeff^\mathrm{BE}}{160^{-\beta_\mathrm{eff,1}}}E(\lambda)
\end{equation}
and 
\begin{equation}
E(\lambda) = \left\{
  \begin{array}{ll}
    \lambda^{-\beta_\mathrm{eff,1}} & \lambda < \lambda_b \\
    (\lambda_b^{\beta_\mathrm{eff,2}-\beta_\mathrm{eff,1}}) \lambda^{-\beta_\mathrm{eff,2}} & \lambda \geq \lambda_b \\
  \end{array}
  \right.,
\end{equation}
where $\lambda_b$ is the break wavelength and is limited to $\geq
175~\micron$.  This emissivity law is similar in form to that used by
\citet{Li01} for astronomical silicates.  The value of
$\kappaeff^\mathrm{BE}$ is set by 
fitting of the diffuse MW SED (\S\ref{sec_kappa_cal} and Table~\ref{tab_mw_diffuse}).

As we are particularly interested in measuring the submm excess, we
define the submm excess as the excess emission at a particular submm
wavelength above or below 
that expected for a SMBB model with $\betaeff = \beta_\mathrm{eff,1}$.
Given the BEMBB model definition, the submm excess at 500~\micron\ is
\begin{equation}
e_\mathrm{500} = \left( \frac{\lambda_b}{500} \right)^{\beta_\mathrm{eff,2} -
  \beta_\mathrm{eff,1}} - 1.
\end{equation}
Using $e_\mathrm{500}$ as one of the fit parameters (instead of
$\beta_\mathrm{eff,2}$), the fit 
parameters for the BEMBB model are 
$\theta_\mathrm{BE} = (\Sigma_d, \dusttemp, \beta_\mathrm{eff,1}, \lambda_b,
e_\mathrm{500})$.   Note that the value of $e_\mathrm{500}$ can be
negative and this would indicate a submm deficit.  The values
for the dust properties are effective values due to composition and temperature
mixing along the line-of-sight and are not directly comparable to
interstellar dust grain analogs studied in the laboratory (see \S\ref{sec_intro}).

\subsection{TTMBB: Two-Temperature Model}

The TTMBB predicts the surface brightness assuming two dust populations
with distinctly different dust temperatures modified by a 
single, non-broken emissivity law.  The surface brightness is then
\begin{equation}
S_{\lambda} = \kappa_\lambda
    \left[ 
     \Sigma_{d1} B_\lambda(T_{\mathrm{eff},d1}) + 
     \Sigma_{d2} B_\lambda(T_{\mathrm{eff},d2}) \right]
\end{equation}
where 
\begin{equation}
\kappa_\lambda = \frac{\kappaeff^\mathrm{TT}}{160^{-\betaeff}} \lambda^{-\betaeff},
\end{equation}
the subscripts $d1$ and $d2$ refer to the two dust components, and
$T_{\mathrm{eff},d1} > T_{\mathrm{eff},d2}$.  
The value of $\kappaeff^\mathrm{TT}$ is set by
fitting of the diffuse MW SED (\S\ref{sec_kappa_cal} and Table~\ref{tab_mw_diffuse}).  

For this model the submm excess at 500~\micron\ is
\begin{equation}
e_\mathrm{500} = \frac{\Sigma_{d2} B_\mathrm{500}(T_{\mathrm{eff},d2})}{\Sigma_{d1}
  B_\mathrm{500}(T_{\mathrm{eff},d1})}.
\end{equation}
Again, we use $e_\mathrm{500}$ as a fit parameter and the full set of fit
parameters for the TTMBB model are
$\theta_\mathrm{TT} = (\Sigma_{d1}, T_{\mathrm{eff},d1}, T_{\mathrm{eff},d2}, \betaeff,
e_\mathrm{500})$.   Note that the value of $e_\mathrm{500}$ for the
TTMBB model \emph{cannot} be
negative unlike the case for the BEMBB model.  The values
for the dust properties are effective values due to composition and temperature
mixing along the line-of-sight and are not directly comparable to
interstellar dust grain analogs studied in the laboratory (see \S\ref{sec_intro}).

\subsection{Restricted $\betaeff$ Models}

It is often assumed in modified blackbody fitting that only $\betaeff$
values between 1 and 2 are valid.  This is based on arguments that
laboratory measurements of dust analogs only give $\beta$ values
between these limits.  More precisely, laboratory measurements of
carbonaceous and silicate dust analogs give $\beta$ values between 0.8
and 2.5 for the Herschel wavelength range \citep[e.g.,][]{Jager98,
Coupeaud11}.  It is clear that luminosity weighted mixing of dust
analogs with $\beta$ values between 0.8 and 2.5 will always result in
$\betaeff$ values in the same range.  Yet this is not necessarily the
case for temperature mixing along the line-of-sight \citep{Juvela12}.
Combining the effects of composition and temperature mixing using full
radiative transfer models, \citet{Ysard12} give evidence that find
that an $\betaeff$ ($\beta_\mathrm{color}$ in their terminology)
between 0.8 and 2.5 is reasonable for a range of realistic cases.
Thus, we include versions of the SMBB, BEMBB, and TTMBB models that
have $\betaeff$ values restricted to be between 0.8 and 2.5.  But we
caution that it is more statistically correct to include $\betaeff$
values outside this range as measurement noise can create SEDs that
require non-physical $\betaeff$ values to provide statistically robust
fits.

\subsection{Band Integration}
\label{sec_band_int}

Our models produce SEDs that are well sampled in wavelength, but our
observations have a very coarse wavelength sampling as they are taken
through filters with broad response functions.  It is important to
correctly model the effects of these broad response functions on the
models to give accurate fits to the observations.  For this paper, we
start with the model predictions of the surface brightnesses at a wavelength 
resolution that well resolves the PACS and SPIRE bandpasses
\citep{Muller11, Griffin13}.  Then, the band surface brightnesses were
determined by integrating over their respective band response functions using
\begin{equation}
S_\mathrm{band} = \frac{ \int S_\nu R_E(\nu) d\nu }
   {\int (\nu_o/\nu)^{-1} R_E(\nu) d\nu }
\label{eq_band_int}
\end{equation}
where $R_E(\lambda)$ is the response function appropriate for extended
sources given in fractional transmitted energy.  The $\nu_o =
c/\lambda_o$ values are given by $\lambda_o = $ 100, 160, 250, 350,
and 500~\micron\ for the bands with the same names.
Eq.~\ref{eq_band_int} mathematically models the data that is produced
by the PACS and SPIRE instruments and data reduction pipelines.  The
integration is done in energy units (e.g., MJy~sr$^{-1}$) as both
instruments use bolometers that measure energy (not photons).  The
denominator of this equation normalizes $R_E(\lambda)$ and accounts
for the PACS and SPIRE calibration convention where the calibration is
given at specific wavelengths ($\lambda_0$) and for a $S(\nu) =
\nu^{-1}$ reference spectrum.

\section{Fitting Technique}

\begin{deluxetable}{lcc}
\tablewidth{0pt}
\tablecaption{Grid Parameters \label{tab_grid}}
\tablehead{\colhead{Parameter} & \colhead{Range} & \colhead{Spacing} }
\startdata
\multicolumn{3}{c}{SMBB} \\ \tableline
$\log(\Sigma_d)$ [$\mathrm{M}_\sun~\mathrm{pc}^{-2}$] & -4 to 1 & 0.1 \\
$\dusttemp$ [K] & 5 to 75 & 1 \\
$\betaeff$ & -1 to 4 & 0.25 \\ \tableline
\multicolumn{3}{c}{BEMBB} \\ \tableline
$\log(\Sigma_d)$ [$\mathrm{M}_\sun~\mathrm{pc}^{-2}$] & -4 to 1 & 0.1 \\
$\dusttemp$ [K] & 5 to 75 & 1 \\
$\beta_\mathrm{eff,1}$ & -1 to 4 & 0.25 \\
$\lambda_b$ [$\micron$] & 175 to 375 & 25 \\
$e_\mathrm{500}$ & -1 to 2 & 0.25 \\ \tableline
\multicolumn{3}{c}{TTMBB} \\ \tableline
$\log(\Sigma_{d1})$ [$\mathrm{M}_\sun~\mathrm{pc}^{-2}$] & -4 to 1 & 0.1 \\
$T_\mathrm{eff,d1}$ [K] & 5 to 75 & 2 \\
$T_\mathrm{eff,d1}$ [K] & 4 to 75 & 2 \\
$\betaeff$ & -1 to 4 & 0.25 \\
$e_\mathrm{500}$ & 0 to 2 & 0.25 
\enddata
\end{deluxetable}
 
We computed the models on discrete grids with spacings fine enough to
resolve the final 1D likelihoods for each parameter.  The grids were
computed over a large range in each parameter to ensure that the
likelihood function was well sampled.  The ranges and spacings for
both models are given in Table~\ref{tab_grid}.  We use a logarithmic
spacing for $\Sigma_d$ to provide a computationally efficient sampling
of the full dynamic range of this parameter.  The minimum and
maximum ranges of the parameters were set iteratively, expanding the
fit parameter ranges until the 1D likelihood function for the vast
majority of the pixels in the galaxies were well sampled.

We fit each pixel that was detected at $3\sigma$ above the background
in all five bands.  The probability that a particular
model fits the data was computed assuming a multi-variate Normal/Gaussian
distribution \citep{Gut09} using
\begin{equation}
  P(\vec{S}^\mathrm{obs} \midline \theta) \, = \, \frac{1}{Q}
  \exp\left(-\frac{1}{2}\chi^2(\theta)\right), 
\label{eq_like1}
\end{equation}
where
\begin{equation}
Q^2 = (2\pi)^n \mathrm{det}\left| \mathbb{C} \right|
\end{equation}
and
\begin{equation}
\chi^2(\theta) = [\vec{S}^\mathrm{obs} - \vec{S}^\mathrm{mod}(\theta)]^T
  \mathbb{C}^{-1}
  [(\vec{S}^\mathrm{obs} - \vec{S}^\mathrm{mod}(\theta)].
\label{eq_chisqr}
\end{equation}
$\vec{S}^\mathrm{obs}$ is the observed SED for a single pixel in the
$n = 5$ bands, $\vec{S}^\mathrm{mod}$ is the SED for a particular
model and parameter set, $\theta$, and $\mathbb{C}$ is the covariance
matrix.  The $^T$ notation denotes the transpose of the vector.  The
covariance matrix is often given as the $\Sigma$ symbol, 
but we have chosen to use $\mathbb{C}$ to avoid confusion with the
dust surface density or standard summation symbol.

The explicit use of a covariance matrix in the fitting allows us to
directly account for correlations between bands in the data.  This is
a different approach than has been recently taken by other authors.
One technique for investigating the effects of correlated noise on
model fit parameters is to perform many Monte Carlo trials of the
observations where they are perturbed by the random and correlated
noise and fit with the model \citep[e.g.,][]{Galliano11}.  A second
technique is to include parameters in a hierarchical Bayesian model
for the correlations in the absolute flux calibration between bands
and then marginalize (integrate) over them to determine their final
fit probabilities \citep[e.g.,][]{Kelly12}.  While not often done, it
is critical to account for the correlated noise in observations as
neglecting such noise terms can significantly bias the resulting fit
parameters \citep{Veneziani13}.  By including the covariance directly
into the likelihood function we do not need to perform many Monte
Carlo trials for every pixel or use a hierarchical Bayesian model to
account for this noise term.  In other words, we can include the
correlations directly in the individual fits efficiently without
having to appeal to the ensemble behavior.

\subsection{LMC and SMC Covariance Matrices}
\label{sec_mc_cov_mat}

For this work, the covariance matrix is defined as 
\begin{equation}
\mathbb{C} = \mathbb{C}_\mathrm{cal} + \mathbb{C}_\mathrm{bkg}
\end{equation}
where $\mathbb{C}_\mathrm{cal}$ is the absolute surface brightness
covariance matrix and $\mathbb{C}_\mathrm{bkg}$ is the background
covariance matrix.  The units of these covariance matrices are
$(\mathrm{MJy/sr})^2$.

The $\mathbb{C}_\mathrm{cal}$ is given by the details of the PACS and
SPIRE absolute flux calibrations.
The SPIRE instrument has been calibrated using a model of Neptune with
an absolute uncertainty correlated between bands for point sources of
4\% and a repeatability that is uncorrelated between bands of 1.5\%
\citep{Griffin13, Bendo13}.  For extended sources, it is recommended
to add an additional 4\% to account for the correlated uncertainty in
the total beam area resulting in an 8\% 
correlated uncertainty between bands \citep{HSO11}.  The PACS
instrument has been calibrated using models of stars and asteroids
with an absolute uncertainty correlated between bands for point
sources of 5\% and a repeatability uncorrelated between bands of 2\%
\citep{Muller11_PACS, Balog13}.  Similar to SPIRE,
for extended sources we add an additional 5\% correlated uncertainty
to account for uncertainties in the total beam area resulting in a
10\% correlated uncertainty between bands.  Finally, we assume the
PACS and SPIRE calibrations are independent given that
PACS is calibrated using stars and SPIRE using Neptune.  Given this
information the elements of $\mathbb{C}_\mathrm{cal}$ are 
\begin{equation}
(\mathbb{C}_\mathrm{cal})_{ij} = S^\mathrm{mod}_i(\theta) S^\mathrm{mod}_j(\theta)
\left[ (\mathbf{A}_\mathrm{cor})_{ij} + (\mathbf{A}_\mathrm{uncor})_{ij} \right]
\end{equation}
where 
\begin{equation}
\mathbf{A}_\mathrm{cor} = 
  \left[ \begin{array}{ccccc}
0.1^2 & 0.1^2 & 0 & 0 & 0 \\
0.1^2 & 0.1^2 & 0 & 0 & 0 \\
0 & 0 & 0.08^2 & 0.08^2 & 0.08^2 \\
0 & 0 & 0.08^2 & 0.08^2 & 0.08^2 \\
0 & 0 & 0.08^2 & 0.08^2 & 0.08^2 \end{array} \right]
\end{equation}
and
\begin{equation}
\mathbf{A}_\mathrm{uncor} = \left[
\begin{array}{ccccc}
0.02^2 & 0 & 0 & 0 & 0 \\
0 & 0.02^2 & 0 & 0 & 0 \\
0 & 0 & 0.015^2 & 0 & 0 \\
0 & 0 & 0 & 0.015^2 & 0 \\
0 & 0 & 0 & 0 & 0.015^2 \end{array} \right].
\end{equation}

The background covariance matrix, $\mathbb{C}_\mathrm{bkg}$ is
calculated empirically from a large set of pixels visually identified
as lying outside of the emitting region of each galaxy.  The
background pixels are in the full images and were processed as
described in \S\ref{sec_data}.  The terms of the
covariance matrix are calculated using
\begin{equation}
\sigma_{ij}^2 = \frac{
  \sum_k^N \left( S_i^k - \left< S_i \right> \right) 
           \left( S_j^k - \left< S_j \right> \right) }
  {N-1} 
\end{equation}
where $N$ is the number of background pixels, $S_i^k$/$S_j^k$ is the
$i$th/$j$th band of the $k$th pixel, and $\left< S_i \right>$/$\left<
S_j \right>$ is the average background in the $i$th/$j$th band.  For
the LMC, $N = 52113$ and 
\begin{equation}
\mathbb{C}_\mathrm{bkg}(LMC) = \left[
\begin{array}{ccccc}
4.23 & 0.78 & 0.65 & 0.33 & 0.14 \\
0.78 & 2.37 & 0.85 & 0.43 & 0.18 \\
0.65 & 0.85 & 0.91 & 0.47 & 0.20 \\
0.33 & 0.43 & 0.47 & 0.25 & 0.11 \\
0.14 & 0.18 & 0.20 & 0.11 & 0.057 \end{array} \right]
\end{equation}
and for the SMC, $N = 4012$ and 
\begin{equation}
\mathbb{C}_\mathrm{bkg}(SMC) = \left[
\begin{array}{ccccc}
2.64 & 0.56 & 0.30 & 0.14 & 0.064 \\
0.56 & 1.18 & 0.46 & 0.23 & 0.094 \\
0.30 & 0.46 & 0.36 & 0.20 & 0.089 \\
0.14 & 0.23 & 0.20 & 0.12 & 0.054 \\
0.064 & 0.094 & 0.089 & 0.054 & 0.030 \end{array} \right] .
\end{equation}
These empirical covariance
matrices illustrate that background is highly correlated with the
correlation increasing in strength towards longer wavelengths.  This
is illustrated by the correlation matrix (terms are
$\mathbb{C}_{ij}/[\sigma_i \sigma_j]$) for the SMC:
\begin{equation}
\mathrm{corr}_\mathrm{bkg}(SMC) = \left[
\begin{array}{ccccc}
1.00 & 0.32 & 0.31 & 0.25 & 0.23 \\
0.31 & 1.00 & 0.70 & 0.61 & 0.49 \\
0.30 & 0.70 & 1.00 & 0.94 & 0.85 \\
0.25 & 0.61 & 0.94 & 1.00 & 0.91 \\
0.23 & 0.49 & 0.85 & 0.91 & 1.00 \end{array} \right] .
\end{equation}
The LMC correlation matrix is very similar and so is not shown.  The
positive and non-zero correlation terms is a signature that the
correlated noise in the background is due to real astronomical
signals.  In this case, it is traceable to the residual foreground MW cirrus
emission and the integrated emission from background galaxies.  The
higher covariance values for the LMC is a reflection of the increased
difficulty of background subtraction for this galaxy.

\subsection{Example Fitting Results}

\begin{figure*}[tbp]
\epsscale{1.15}
\plotone{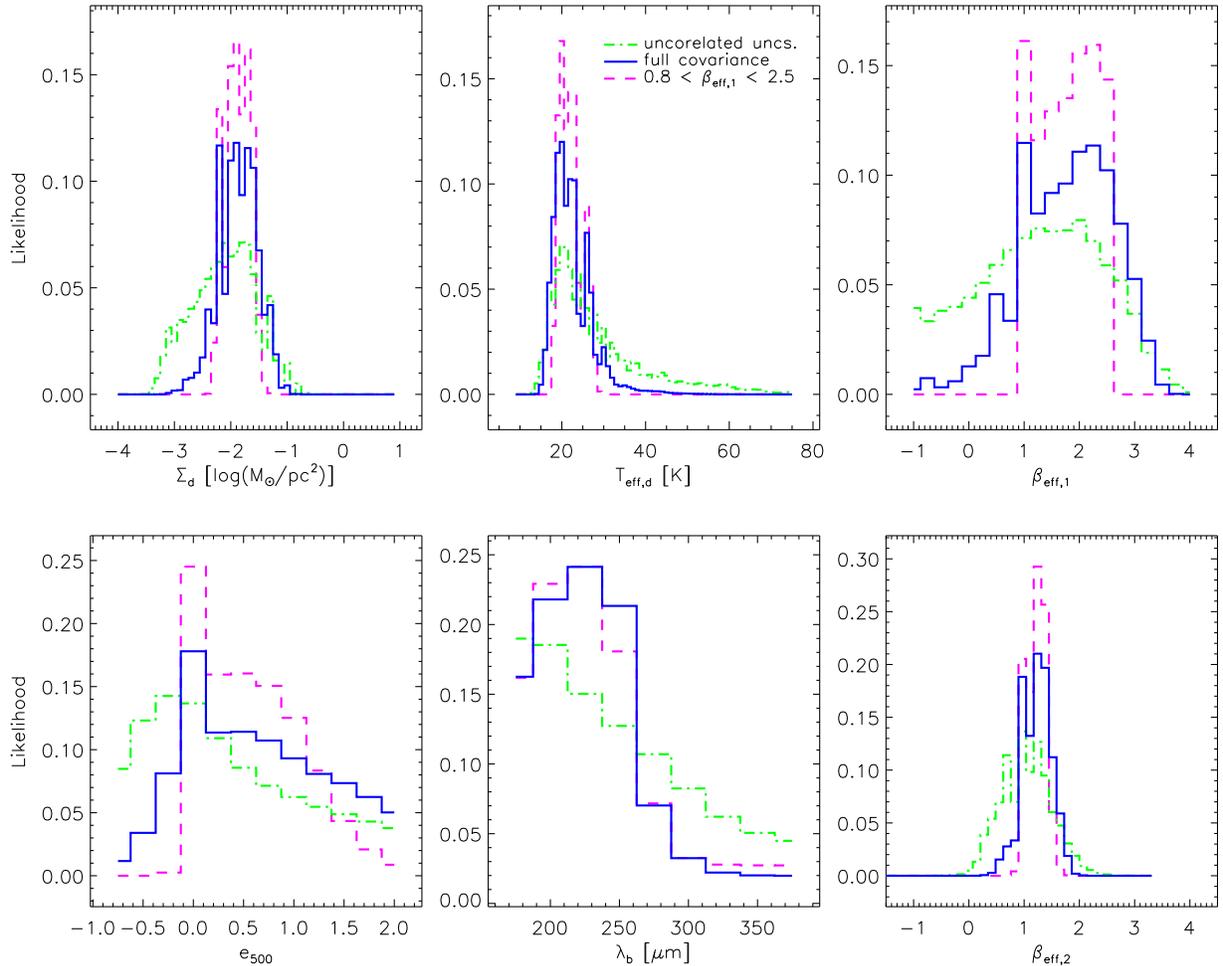}
\caption{The 1D likelihood functions for a single pixel in the SMC
  using the BEMBB 
  model are plotted for fitting while assuming uncorrelated uncertainties,
  including the full covariance, and including the 
  full covariance while restricting the allowed $\beta_\mathrm{eff,1}$ values to be between
  0.8 and 2.5.  Note that $\beta_\mathrm{eff,2}$ is completely
  determined by the value of $\beta_\mathrm{eff,1}$ and $e_{500}$ and we present the
  $\beta_\mathrm{eff,2}$ 1D likelihood function for completeness.
  \label{fig_cov_impact}}
\end{figure*}

The fitting technique we use fully computes the $n$D likelihood function
that a particular model fits the SED of a pixel where $n$ is the
number of fit parameters.  One way to visualize the results is to
create 1D likelihood functions for each fit parameter by marginalizing
(integrating) over all the other parameters.  This is shown in
Fig.~\ref{fig_cov_impact} for the BEMBB model for a single pixel in
the SMC for three different
assumptions; assuming uncorrelated uncertainties, including the full
covariance, and including the full covariance while restricting $\beta_\mathrm{eff,1}$ to
vary between 0.8 and 2.5.  The results for pixels in the LMC are
similar.  With the same overall uncertainties, we 
obtain a much narrower function with a stronger likelihood by including the known
covariance between the bands (\S\ref{sec_mc_cov_mat}) than by assuming
that there is no correlation between bands.  In this case, including
the known covariance between bands results in better constraints on
the fit parameters as the allowed model space is reduced.
The impact of a limited $\beta_\mathrm{eff,1}$ range is shown in this
figure where, not surprisingly, it makes for a narrower 1D likelihood
function than allowing $\beta_\mathrm{eff,1}$ to vary to fully sample
the $\beta_\mathrm{eff,1}$ 1D likelihood function.  Note that this
limitation simply crops the $\beta_\mathrm{eff,1}$ 1D likelihood
function, but changes the shape of the other 1D likelihood functions
significantly.

\subsection{Sensitivity Tests}
\label{sec_sense_tests}

\begin{figure*}[tbp]
\epsscale{1.1}
\plotone{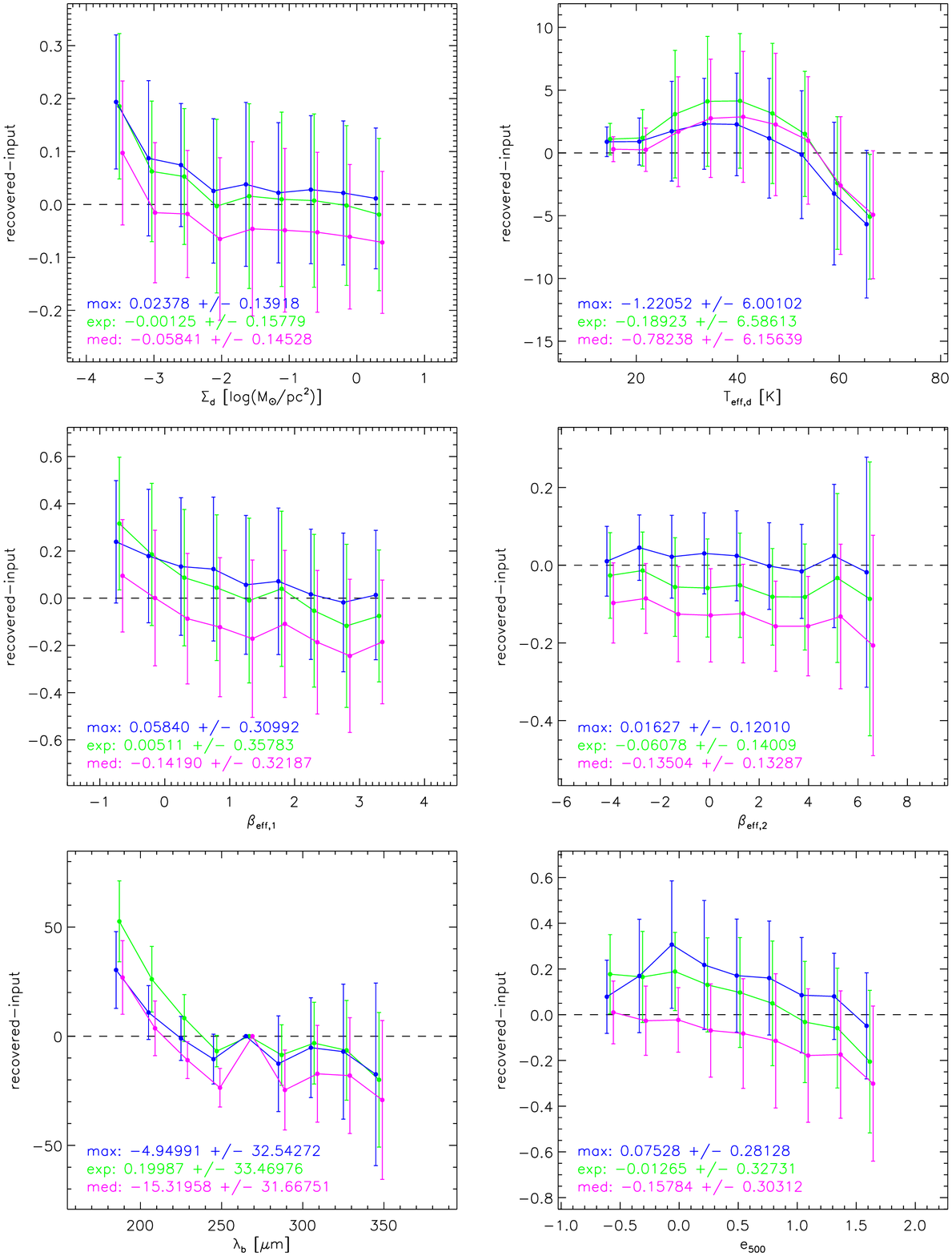}
\caption{The results for sensitivity tests of the BEMBB model for 2000
  models randomly selected from the full model grid 
  are shown.  The results are plotted 
  as averages and standard deviations of the recovered minus input
  parameters in 10 bins over the parameter range.  The three
  different methods of determining the accuracy of the recovered parameters are
  'max' = maximum likelihood, 'exp' = expectation value, and 'realize' = one
  realization based on the 1D likelihood functions for each parameter.
  \label{fig_sens_tests}}
\end{figure*}

The goal of the sensitivity tests is to determine if there are
systematic shifts in recovered parameters and if the uncertainty on
the recovered parameters matches that measured from the widths of the
1D likelihood functions.  We simulated observations by picking a model
SED and adding noise using the Cholesky factorization of the
covariance matrix appropriate as if the model was
observed like the SMC was observed in HERITAGE.  The results using the
LMC noise model give very similar results.  We repeated the
simulation for each model SED 20 times to provide a good sampling of
the recovered fit parameter uncertainties and systematic offset from
the input fit parameters.  

As we are testing the ability of this fitting technique to recover
parameters by fitting simulated observations, this requires a way to
measure the recovery of the input model parameters.  The main output
of the fitting is the $n$D likelihood function, but it is often useful
to distill these results to ``best fit'' or summary values.  We use
three different ways to define the ``best fit'' values.  The first is
the most traditional definition of the ``best fit'' and corresponds to
the maximum likelihood (`max').  This is also called the ``traditional
$\chi^2$'' method in some papers \citep[e.g.,][]{Kelly12, Juvela13}.
The `max' value is most useful when plotting the best fitting model
with observations or investigating the fitting residuals.  The second
is the expectation value (`exp') which is the likelihood weighted
average of the parameter and is a reflection of the full likelihood
function.  This `exp' value reflects the best ``average'' value as it
reflects the full likelihood function (not just the peak like the
`max' value).  We find the `exp' particularly useful for making images
of the fit parameters.  The third way to reflect the best fit is take
a realization of the full $n$D likelihood function itself (`realize').  This
involves randomly sampling the likelihood function and reflects the
full likelihood function's shape in a statistical sense.  The
`realize' method is most useful when studying the ensemble behavior of
the fit parameters for many pixels.

The results for runs with 2000 randomly picked BEMBB models are shown
in Fig.~\ref{fig_sens_tests}.  All three different methods of
determining the ``best fit'' parameters give similar results with
similar trends with each parameter.  The `exp' gives the lowest
systematic error in the recovery, but the `max' gives the lowest
scatter.  The `realize' method provides a nominally worse recovery
than both the other methods, but is a fuller picture of true
sensitivity of the fitting.  Overall, which ``best fit'' method used
depends on the particular question being asked.  We illustrate
this later in this paper and in the companion paper on the gas-to-dust
ratio (Roman-Duval et. al., this issue).

Of particular interest for this paper is the fact that the recovery of the submm
excess, $e_\mathrm{500}$, is good to around 10\%, on average, for
the `realize' method and around 1\% for the `exp' method.
For the companion paper (Roman-Duval et al, this issue), the fit parameter of
main interest is $\Sigma_d$ and the recovery is good, on average and in
$\log(\Sigma_d)$ units, to 0.05 for the `realize' method and 0.001 for the `exp'
method.  This excellent recovery of $\log(\Sigma_d)$ holds even in the 
presence of significant 
scatter in $\dusttemp$ and may be due to other parameters in the fitting varying to
compensate.  Note that for the `exp' method we computed the
expectation value of $\log(\Sigma_d)$ as we found that the sensitivity
tests showed significantly less systematic bias than if we computed
the expectation value of $\Sigma_d$.  We confirmed that the widths of
the 1D likelihood functions matches the noise in the recovery of the
input model parameters.

\subsection{Number of Parameters and Data Points}

The number of parameters in our models is three, five, and five for
the SMBB, BEMBB, and TTMBB models, respectively.  In this paper, these
models are fit to FIR-submm SEDs that are composed of five data
points.  At first glance, this violates the rule that fitting requires
at least one data point more than the number of fitting parameters to
provide a unique solution.  This is correct, if the fitting is done
with a model that can fit any distribution of data points.  This is
clearly not the case for our models as they are all constrained to
have a spectral shape of one or two modified blackbodies.  In other
words, they cannot fit arbitrary spectral shapes but are constrained
by our knowledge of the physics of dust grain emission.  Effectively,
we are using more than just five data points in our fits as we combine
the data points with a larger body of observations that informs our
understanding of dust physics and, therefore, the appropriate models
to use.  Finally, our use of full likelihood functions explicitly
accounts for the impact of the number of parameters on how well we can
determine each fit parameter.  Using full likelihood functions has the
additional benefit of measuring how well each parameter is constrained
by the data explicitly. Some parameters are better constrained than
others as shown in Fig.~\ref{fig_cov_impact}.  For example, $\Sigma_d$
and $\dusttemp$ are better constrained as the
overall level and spectral shape are well constrained by the
observations, but the detailed spectral shape is less well constrained
and this impacts $\beta_\mathrm{eff,1}$, $\lambda_b$, and
$e_{500}$ strongly.

\section{Model Calibration}
\label{sec_kappa_cal}

It is important to calibrate dust models to reproduce observations
where there are independent measurements of the same quantities using
the same fitting technique.  This is regularly done when setting up
full dust grain models \citep[e.g.,][]{Li01, Zubko04, Compiegne11}.
One key calibration source is the FIR--submm SED of the MW diffuse
ISM.  This is a unique environment as it is the one place where the
amount of dust has been measured using ultraviolet and optical
gas-phase absorption lines and knowledge of the total amount of atoms
expected in the ISM \citep[e.g.,][]{Jenkins09}.  Thus, fitting the
FIR-submm MW diffuse SED results in a calibration of the dust
emissivity $\kappa_\lambda$ as the degeneracy between this quantity
and $\Sigma_d$ is removed.

In full dust grain models, the calibration of $\kappa_\lambda$ is
usually set such that the luminosity weighted average response of the
different dust grain components reproduces the MW diffuse SED when the
dust is illuminated by the average MW radiation field.  In a similar
manner, the $\kappaeff$ for the models used in this paper is set such
that fitting the MW diffuse SED produces the observed gas-to-dust
ratio.  By determining $\kappaeff$ using the measurements of the
diffuse MW emission for each of our models, we ensure that our models
derive the correct dust surface density in the one physical
environment where we have independent constraints on the dust mass.
It is critical to note that this calibration does not impose a
gas-to-dust ratio calibration on our model, just a calibration that we
derive the correct mass of dust in the MW diffuse ISM.

This calibration does mean that we are assuming that the dust
properties in the Magellanic Clouds are the same as those in the
diffuse MW.  This assumption is reasonable given the evidence from
ultraviolet extinction measurements in all three galaxies.  The SMC
does show UV extinction curves most different from the average in
the MW, but it also has curves that are very similar to the MW average
\citep{Gordon98, MaizAppellaniz12}.  The LMC shows extinction curves
that are similar or equivalent to the MW average \citep{Misselt99,
Gordon03}.  While many of the MW lines-of-sight show extinction curves
similar to the MW average by definition \citep{Valencic04}, there is
one line-of-sight that shows a UV extinction curve indistinguishable
from the most different SMC extinction curves \citep{Valencic03}.  It
is not clear if the globally average UV dust extinction is different
between the three galaxies, mainly due to small samples sizes of such
measurements in the Magellanic Clouds \citep{Gordon03}.  One piece of
evidence that far-IR emissivity of dust grains is similar between the
MW and SMC is the similarity of their $\kappaeff$ values as derived
using dust grain model fitting (see \S\ref{sec_cal_kappa}).  While it is
reasonable to assume the dust is similar in all three galaxies, it is
an assumption and the dust surface densities will vary inversely in
direct proportion to any changes in the adopted $\kappaeff$
calibration.  

Evidence for different than the MW dust in the LMC was found in work
by \citet{Meixner10} and \citet{Galliano11} using the HERITAGE test
observations of a strip in the LMC.  These works used two models of
dust, one composed of silicates, graphite, and PAH grains that
describes average MW dust (``standard'') and a second with amorphous
carbon instead of graphite (``AC'').  The analysis found that the
gas-to-dust ratio for the ``standard'' model was lower than reasonable
for the LMC metallicity, while the ``AC'' model produced a reasonable
ratio.  We discuss the issue of gas-to-dust ratios for the LMC and SMC
using the fitting results for the models used in this paper and
calibrated using the MW diffuse SED in \S\ref{sec_gas_to_dust}.  In
addition, we have estimated the systematic error on $\kappaeff$ due to
assuming that the dust is like that in the MW in \S\ref{sec_cal_kappa}.

Direct measurements of ISM depletions in the Magellanic Clouds would
allow us to directly calibrate our models in these galaxies.  This
would remove the assumption that the dust grain compositions in the
Magellanic Clouds are the same as those in the Milky Way.  Currently,
there exists only a limited number of sightlines and atoms with
measured depletions in the Magellanic Clouds \citep{Roth97, Welty97, Welty01,
  Sofia06, Peimbert10, Welty10}. Extending these studies in terms of 
atomic species and galactic environments should be a priority for the
astronomical community, since they are critical for interpreting the
wealth of FIR to submm ISM observations obtained by recent space
missions. 

\subsection{Milky Way Diffuse SED}

\begin{figure}[tbp]
\epsscale{1.2}
\plotone{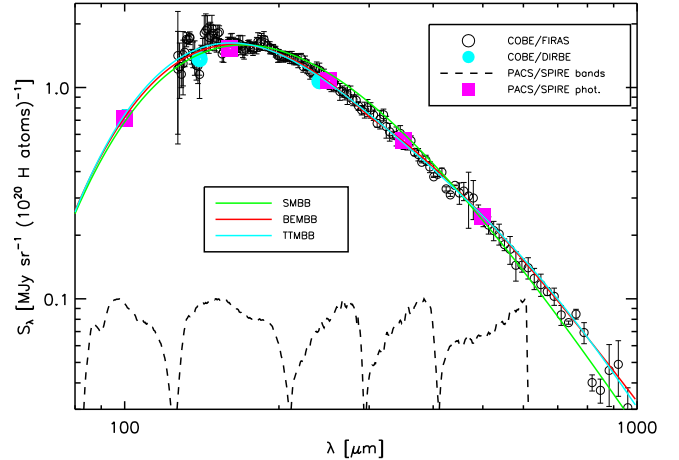}
\caption{The observed MW diffuse SED from COBE FIRAS and DIRBE is
  plotted along with the 
  best fits for the models used in this paper.  The best fit is
  defined using the `max' method discussed in \S\ref{sec_sense_tests}.
  The 'PACS/SPIRE 
  phot.' points (purple squares) are those used to constrain the fits of the
  models and were derived from the COBE FIRAS and DIRBE measurements.
  \label{fig_mw_sed}}
\end{figure}

For the diffuse MW emission, we use the \citet{Compiegne11}
measurement where emission was measured by correlating the IR versus
HI emission maps in atomic gas dominated regions of the MW.  The IR
measurements we use are mainly the COBE/FIRAS spectrophotometry from
127 to 1200~\micron\ supplemented by the DIRBE 100~\micron\
photometry.  As we want to calibrate our models using the same bands
as used for the HERITAGE observations, we integrated this diffuse MW
SED using the method described in \ref{sec_band_int} for all the bands
except the PACS 100~\micron\ band.  For this band, we adopted the
DIRBE 100~\micron\ measurement as the bandpasses are similar.  The
resulting MW diffuse SED is 0.71, 1.53, 1.08, 0.56, and 0.25
MJy~sr$^{-1}$~($10^{20}$~H~atom)$^{-1}$ for the 100, 160, 250, 350,
and 500~\micron\ and is plotted in Fig.~\ref{fig_mw_sed}.  These
values differ from those given for the same bands by
\citet{Compiegne11} mostly as we have not included the 0.77 correction
for ionized gas.  In addition, there are minor differences in the
response curves used.  We do not include the 0.77 correction for
ionized gas as the depletion measurements do not include any ionized
gas correction.  For the uncertainties, we have assumed a 5\%
correlated and a 2.5\% uncorrelated terms (see \S\ref{sec_band_int})
given the high quality of the COBE FIRAS and DIRBE calibrations.

\subsection{Milky Way Diffuse Gas-to-Dust Ratio}

As the MW diffuse SED is measured as a correlation between dust and
gas emission, the constraint we need is the MW diffuse gas-to-dust
ratio.  We use the work of \citet{Jenkins09} to determine the
appropriate gas-to-dust ratio since this work provides an excellent
compilation and summary of MW depletions.  The observed H columns of
our adopted FIR-submm MW diffuse SED are $\log[N(H)] < 20.7$.  The
average depletion of all the sightlines with these column densities
tabulated by \citet{Jenkins09} is $F_* = 0.36$.  $F_*$ is the
depletion factor and measures the overall depletions in a sightline.
Using the depletion fits of \citet{Jenkins09} with $F_* = 0.36$, the
diffuse MW gas-to-dust ratio is computed to be 150. 

\subsection{Calibrating $\kappaeff$}
\label{sec_cal_kappa}

\begin{deluxetable*}{lccc}
\tablewidth{0pt}
\tablecaption{MW Diffuse Fit Results \label{tab_mw_diffuse}}
\tablehead{\colhead{Model} & \colhead{$\kappaeff$\tablenotemark{a}} & \colhead{Other
    Parameters} & \colhead{Expectation Values} \\ & [cm$^{2}$~g$^{-1}$] & & }
\startdata
SMBB & $9.6 \pm 0.4 \pm 2.5$ & $(\dusttemp, \betaeff)$  & $(17.2 \pm
0.4~\mathrm{K}, 1.96 \pm 0.10)$ \\
BEMBB & $11.6 \pm 1.5 \pm 2.5$ & $(\dusttemp, \beta_{\mathrm{eff},1}, 
  \lambda_b, e_\mathrm{500})$ & $(16.8 \pm 0.6~\mathrm{K}, 2.27 \pm 0.15,
  294 \pm 29~\micron, 0.48 \pm 0.11)$ \\ 
TTMBB & $517 \pm 214 \pm 2.5$ & $(T_{\mathrm{eff},d1}, T_{\mathrm{eff},d2}, 
  \betaeff, e_\mathrm{500})$ & $(15.0 \pm 0.7~\mathrm{K},
  6.0 \pm 0.8~\mathrm{K}, 2.9 \pm 0.1, 0.91 \pm 0.25)$ \\
TTMBB & $9.6 \pm 0.4 \pm 2.5$ & adopted &
\enddata
\tablenotetext{a}{The results are given as value $\pm$ fitting
  uncertainty $\pm$ systematic uncertainty}
\end{deluxetable*}

We calibrate the value of $\kappaeff$ in each of our models so
that they reproduce the MW diffuse observed gas-to-dust ratio of 150.
For our work, we have chosen 160~\micron\ to set our normalization of
$\kappa_{\mathrm{eff},\lambda}$ as shorter wavelengths have a weaker
dependence on temperature based on laboratory investigations of dust analogs
\citep{Coupeaud11}.  The $\kappaeff$ values required for each model
based on the `exp' method of determining the best fits (see
\S\ref{sec_sense_tests}) are given in Table~\ref{tab_mw_diffuse}.  The
second uncertainty on $\kappaeff$ is an estimate of the systematic
uncertainty (see next paragraph).  The
fit parameters for each model are also given in this table, along with
1$\sigma$ uncertainties.  The larger relative uncertainties on
$\kappaeff$ for the BEMBB model as compared to the SMBB can be
directly traced to the larger number of BEMBB fit parameters.
The `max' best fit models are plotted in
Fig.~\ref{fig_mw_sed}.

The $\kappaeff$ values for the SMBB and BEMBB models agree favorably
with other determinations while the value for the TTMBB model does
not.  For example, if ``astronomical'' silicate grains with $a =
0.1~\micron$ and $\rho = 3$~g~cm$^{-3}$ are used, then $\kappaeff =
13.75$~cm$^{2}$~g$^{-1}$.  Such grain properties are often assumed for
simple modified blackbody fits as this is the average size for a
\citet{Mathis77} grain size distribution \citep{Hildebrand83}.  The
widely used \citet{Weingartner01} full dust grain model for R(V)~=~3.1
has a $\kappaeff = 9.97$~cm$^{2}$~g$^{-1}$.  The updated version of
this model has a $\kappaeff = 12.5$~cm$^{2}$~g$^{-1}$
\citep{Draine07model, Draine14}.  The $\kappaeff$ values for the \citet{Zubko04}
models that include graphite and amorphous carbon range from 10.75 to
15.0~cm$^{2}$~g$^{-1}$.  Finally, the \citet{Weingartner01} model for
the SMC Bar extinction curve with no 2175~\AA\ extinction feature has
$\kappaeff = 13.1$~cm$^{2}$~g$^{-1}$.  Using the range of these model
$\kappaeff$ values we estimate that there is a $\pm
2.5$~cm$^{2}$~g$^{-1}$ additional uncertainty on $\kappaeff$ due to
systematic uncertainties in our knowledge of dust grains.

The TTMBB model with $\kappaeff = 517 \pm 214$~cm$^{2}$~g$^{-1}$
requires a dust grain 
that is very efficient at emission, yet this level of efficiency is
much higher than any astronomically reasonable dust grain.  A much
simpler explanation is that the dust in the MW diffuse ISM is not well
modeled by a TTMBB model that includes a very cold ($\dusttemp \sim
6~\mathrm{K}$) dust grain population.  This is the same conclusion
given by the \citet{Reach95} analysis of the FIRAS data.  There still
may be regions in the ISM of the MW or other galaxies that are well
described by the TTMBB model.  To allow for such regions, we adopt the
$\kappaeff$ of the SMBB model as the value for the TTMBB model.

The variations in the $\kappaeff$ values in the literature and between
the different models used in this paper clearly indicate that
$\kappaeff$ is sensitive to the model assumptions.  Thus, it is
important to calibrate each model explicitly with the diffuse MW SED
and a depletion measured gas-to-dust ratio.  This is a standard
calibration method for dust grain models \citep{Draine07model,
Compiegne11} and we advocate that such calibrations be done for all
dust emission models \citep{Bianchi13}.  Such model calibrations will
allow for meaningful comparisons between the results from different
models.

\section{Results}
\label{sec_results}

\subsection{Fitting Residuals}
\label{sec_fit_residuals}

\begin{figure*}[tbp]
\epsscale{0.9}
\plotone{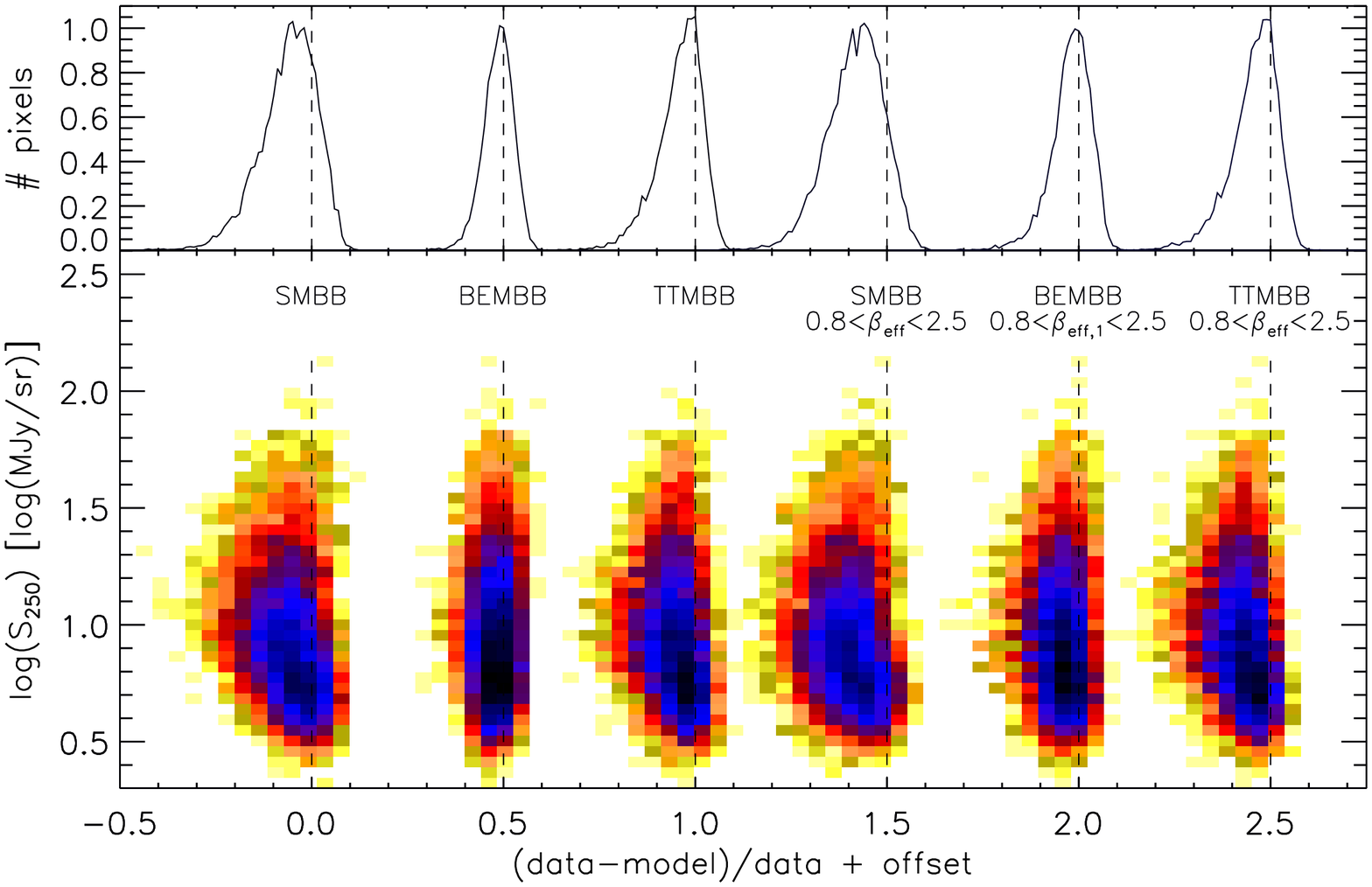}\\
\plotone{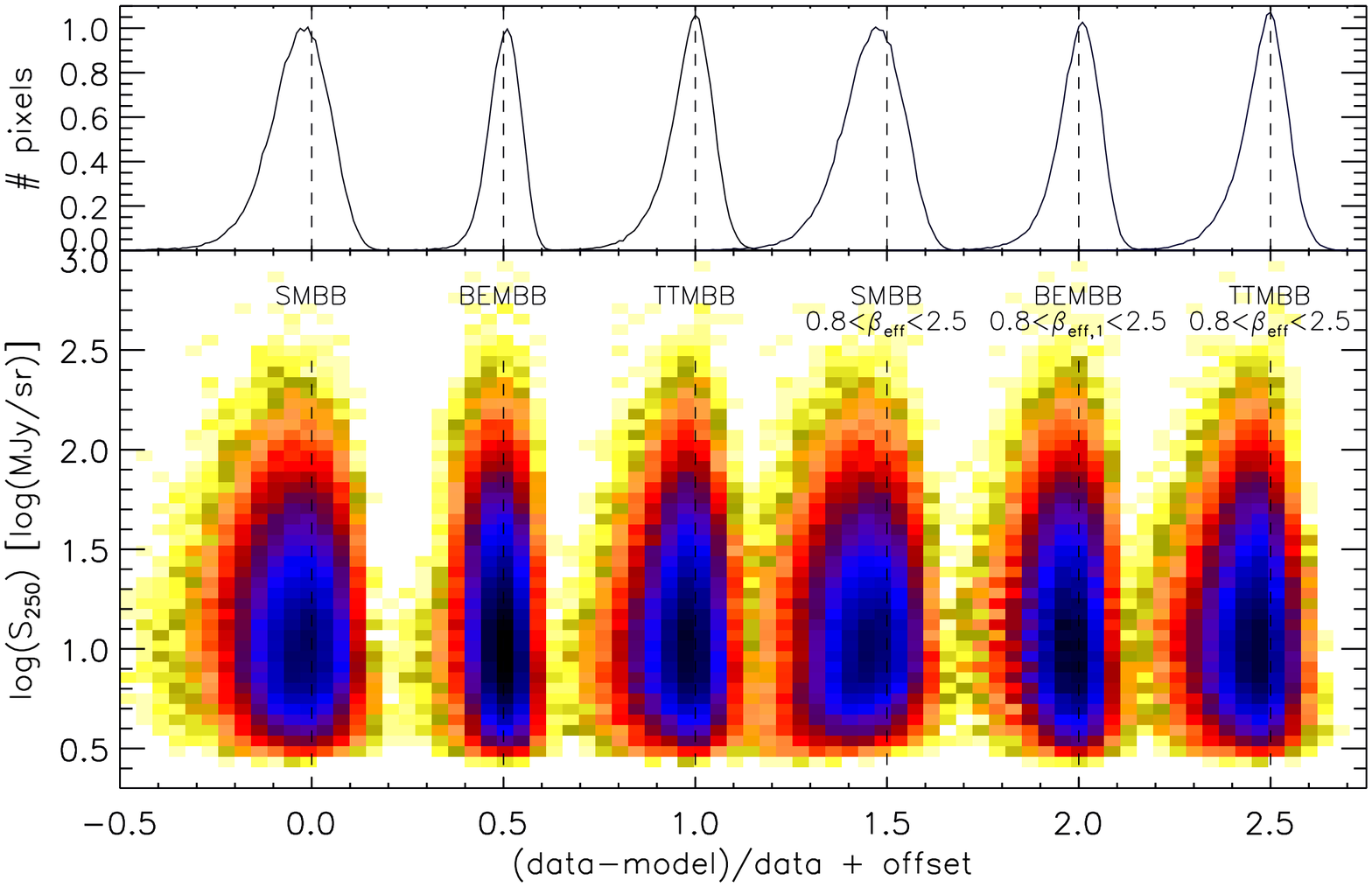}
\caption{The fractional residuals for the SMC (top) and LMC (bottom)
  of the fits for the SPIRE 
  250~\micron\ band are shown for all the models.  Each model
  has been plotted shifted by multiples of 0.5 on the x-axis.  The
  false color gives the log density of points and each point
  represents the residual for the `max' estimator for a single pixel.  The `max'
  estimator was used to give each model the best chance to have the
  lowest residuals. The plots at other wavelengths show similar
  behaviors with the BEMBB model having the lowest residuals.
  \label{fig_fit_resid}}
\end{figure*}

One obvious question is: Which model, SMBB, BEMBB, or TTMBB, fits the
observations best?  The answer to this question will give an
indication of the origin of the submm excess.  The most
straightforward method to test how well a model fits the data is to
examine the residuals of the data to the fits.  The $\chi^2$ value
computed using eq.~\ref{eq_chisqr} gives such a quantitative measure
of the residuals.  For the SMC, the pixel averaged $\chi^2$ value is
3.47 for the SMBB model, 0.88 for the BEMBB model, and 1.83 for the
TTMBB.  The models with $0.8 < \betaeff < 2.5$ have higher average
$\chi^2$ values than the unconstrained versions.  For example, the
$\betaeff$ constrained version of the BEMBB model for the SMC has an
average $\chi^2$ value of 1.32.  The LMC average $\chi^2$ values
behave similarly.

More evidence that the BEMBB fits the data best (out of the three
models) can be found by 
examining the behavior of the fit residuals versus surface brightness. 
Fig.~\ref{fig_fit_resid} shows the fit residuals for the SPIRE
250~\micron\ band for all three models used in
this paper for both Magellanic Clouds.  The trends for other bands are
similar, especially in the relative behavior of the fit residuals
between the models. 
This figure clearly shows that the simplest model (SMBB) has residuals
larger than expected given the known uncertainties.  This holds for
$\betaeff$ unconstrained and constrained to be between 0.8 and 2.5.
In addition, the residuals for the SMC have a systematic trend with
more negative residuals at intermediate surface brightnesses.  Such a
trend is not consistent with the uncertainties in the absolute flux
calibration or the background subtraction.  Of all models, the BEMBB
model without any constraint on $\betaeff$ fits the data best.
Overall, the BEMBB model shows the smallest residuals with no obvious
trend with surface brightness unlike the other models.  The BEMBB
model consistently shows smaller residuals in all the bands, not just
the SPIRE~250~\micron\ band.  The other models have higher overall
residuals and show systematic offsets and/or trends with surface
brightness.  The BEMBB and TTMBB models have the same number of fit
parameters, yet the behavior of their residuals are different.
This illustrates that it is not only the number of fit parameters that is
critical for the fitting accuracy, but the allowed spectral shapes.

Overall, the BEMBB spectral shapes 
fit the data better than the TTMBB and SMBB spectral shapes.  This is
evidence that the submm excess is more likely to be 
due to emissivity variations than a second population of cold dust.

\subsection{Total Dust Masses}

\begin{deluxetable}{lcc}
\tablewidth{0pt}
\tablecaption{Integrated Dust Masses and Gas-to-Dust Ratios \\
Integrated over $>$$3\sigma$ Pixels \label{tab_dmasses}}
\tablehead{\colhead{Model} & 
  \colhead{$M_d$ [M$_\sun$]} & \colhead{Gas/Dust\tablenotemark{a}}}
\startdata
\multicolumn{3}{c}{LMC} \\ \hline
SMBB & $(8.1 \pm 0.07 \pm 2.1) \times 10^{5}$ & $340 \pm 90$ \\
BEMBB\tablenotemark{b} & $(6.7 \pm 0.03 \pm 1.7) \times 10^{5}$ & $400 \pm 100$ \\
TTMBB & $(1.2 \pm 0.01 \pm 0.3) \times 10^{7}$ & $22 \pm 6$ \\ 
\multicolumn{2}{l}{expected: scaling MW gas-to-dust ratios} & 200-500 \\
\multicolumn{2}{l}{expected: MW depletions and LMC abundances} & 150-360 \\
\multicolumn{2}{l}{expected: all metals in dust} & $\geq$105 \\
\hline
\multicolumn{3}{c}{SMC} \\ \hline
SMBB & $(8.1 \pm 0.1 \pm 2.1) \times 10^{4}$ & $1440 \pm 380$ \\
BEMBB\tablenotemark{b} & $(6.7 \pm 0.1 \pm 1.7) \times 10^{4}$ & $1740 \pm 440$ \\
TTMBB & $(5.1 \pm 0.3 \pm 1.3) \times 10^{5}$ & $230 \pm 60$ \\
\multicolumn{2}{l}{expected: scaling MW gas-to-dust ratios} & 500-1250 \\
\multicolumn{2}{l}{expected: MW depletions and SMC abundances} & 540-1300 \\
\multicolumn{2}{l}{expected: all metals in dust} & $\geq$300 
\enddata
\tablenotetext{a}{The integrated gas masses in M$_\sun$ for the same
  areas and with the same background removal in the
  LMC/SMC are $2.5 \times 10^8$/$1.0 \times 10^8$ for HI and $2.1
  \times 10^7$/$1.6 \times 10^7$ for H$_2$ \citep{Leroy07SMC, Hughes10}.}
\tablenotetext{b}{Model favored from the analysis in this paper (see
  \S\ref{sec_fit_residuals} and \S\ref{sec_gas_to_dust})}.
\end{deluxetable}

The total dust masses are of interest for studies
of the lifecycle of dust in the LMC and SMC \citep{Boyer12, Matsurra13,
  Zhukovska13}.  In addition, they can be used along with the total
gas masses as a way to tell if a model produces realistic amounts
of dust (see \S\ref{sec_gas_to_dust}).

We give the dust masses for the different models in
Table~\ref{tab_dmasses} integrated over the $>$$3\sigma$ pixels.  The
restricted $\betaeff$ version of the 
models produces results that are very similar and are not given in
the table.  The dust mass values are given as total $\pm$ statistical
uncertainty $\pm$ uncertainty due to the $\kappaeff$ uncertainty.  To
convert from dust surface density to dust mass we use distances of
60~kpc \citep{Hilditch05} and 50~kpc \citep{Walker12} for the SMC and
LMC, respectively.  The total dust masses are computed from the `realize'
method to produce dust surface density maps that provide a full
accounting of the likelihood functions for all pixels.  Ten different
maps were made for each galaxy using the `realize' method that samples
the likelihood function once for each pixel.  This provides a robust
measurement of the impact of the fitting noise of each pixel in the
integrated dust mass measurement.  The average and
statistical uncertainty of the integrated dust mass were computed from
the ten maps.  The large
number of pixels in each galaxy results in the total dust mass
changing only slightly between different realizations and this is the
origin of the small statistical uncertainty.  These dust masses are
integrated only over the areas that were detected at $3\sigma$ above
the background in all five {\it Herschel} bands measured by HERITAGE.
Pixels above $>$$3\sigma$ contribute 0.79, 0.73, 0.62, 0.61, and 0.61
of the SMC 
global fluxes of 15.7, 20.8, 14.5, 8.3, and 3.9 kJy for the PACS100,
PACS160, SPIRE250, SPIRE350, and SPIRE500, respectively.  For the LMC,
these fractions are 0.91, 0.89, 0.87, 0.87, and 0.87 for global fluxes
of 223, 259, 142, 73, and 31 kJy for the same bands.  The global
fluxes quoted here differ from those given by \citet{Meixner13} due to our
subtraction of MW cirrus foreground and the additional background
subtraction step.

The quantitative impact of correctly including the correlated
noise in the measurements can be illustrated by noting that assuming
the noise is uncorrelated between bands results in the BEMBB model
giving fits with a total SMC dust mass that is $\sim$50\% higher than
the total dust mass given in Table~\ref{tab_dmasses}.  The importance
of accounting for the full likelihood function is equally important:
the total SMC dust mass for the BEMBB model is $\sim$50\% higher
using the `max' values and $\sim$30\% lower using the `exp' values of
$\log(\Sigma_d)$ when compared to the `realize' value given in
Table~\ref{tab_dmasses}.  The `realize' values are the correct values
for determining the total dust mass values as they statistically
reflect each pixel's full likelihood function, asymmetries and all, in
the sum of the individual pixel masses.  
The `max' and `exp' values only reflect a limited portion of the
likelihood function and this systematically biases the results.  This
is additional evidence that the likelihood functions for $\Sigma_d$
are not well behaved Gaussians centered on the `max' value (see
Fig.~\ref{fig_cov_impact}).

Our total dust masses are only lower limits as we do not include the
dust responsible for the emission with surface brightnesses below
3$\sigma$ in any band. 
We can estimate the dust mass due to these $<$$3\sigma$ regions by
modeling the integrated flux of these regions for each galaxy.
Basically, we fit the 
SED that is the difference from the global fluxes quoted above and the
integrated fluxes from $<$3$\sigma$ pixels.  The resulting integrated
dust masses for the BEMBB model and the $<$$3\sigma$ pixels are $(5.9
\pm 3.6) \times 10^4$ and 
$(1.6 \pm 1.3) \times 10^4$~$\mathrm{M}_\sun$ for the LMC and SMC,
respectively.  The uncertainties are quite large due to the low surface
brightnesses and strong mixing of environments in these integrated SEDs.
Combining the $<$$3\sigma$ pixel dust masses with those for
$>$$3\sigma$ pixel (Table~\ref{tab_dmasses}), we find total dust
masses of $(7.3 \pm 1.7) \times 10^5$ and
$(8.3 \pm 2.1) \times 10^4$~$\mathrm{M}_\sun$ for the LMC and SMC,
respectively.   For reference, the total gas masses that correspond to
the same areas and same background removal as these total dust masses
are $3.1\times 10^8$ and 
$3.0\times 10^8$~$\mathrm{M}_\sun$ for the LMC and SMC, respectively.

\citet{Bot10} obtained global dust masses for both galaxies by fitting
\citet{Draine07} dust models to their global fluxes.  They found
masses of $3.6 \times 10^6$ and $0.29-1.1 \times
10^6$~$\mathrm{M}_\sun$ for the LMC and SMC, respectively.
\citet{Leroy07} fit the spatially resolved Spitzer observations with
\citep{Dale02} models and find a total SMC dust mass of $3 \times
10^5$~$\mathrm{M}_\sun$.  These values of the dust masses are factors
of 4--5 larger than our values. The differences are likely due to
different assumptions in the models used, the fitting techniques, the
broader wavelength range of data, and/or the increased mixing of
environments.

\subsection{Total Gas-to-Dust Ratios}
\label{sec_gas_to_dust}

\begin{figure}[tbp]
\epsscale{1.2}
\plotone{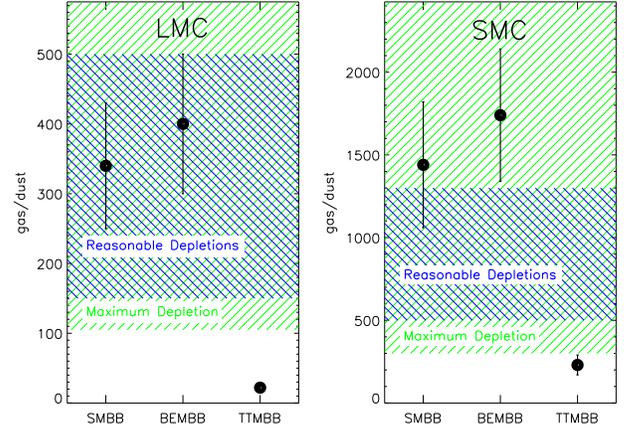}
\caption{The gas-to-dust ratios (GDRs) are plotted as black circles for
  each of the three
  models and for both galaxies.  The ``reasonable'' GDR range expected from
  scaling the MW diffuse to dense GDRs is given as a blue
  hatched region.  The GDR range allowed by assuming the ``maximum''
  depletions is given as a green hatched region (e.g. a lower limit
  on the GDR).
  \label{fig_gdrs}}
\end{figure}

One test of the submm excess origin is to investigate how the
overall gas-to-dust ratios for each model compare to the expected ratios.  
We explore overall gas-to-dust ratios as a test of the 
consistency of each dust model with expectations based on the
measured gas masses and metallicities of the LMC and SMC.  The
detailed spatial behavior of the gas-to-dust ratio with environment is
investigated in Roman-Duval et al. (this issue). 

The gas-to-dust ratios for each galaxy and all three models are given
in Table~\ref{tab_dmasses}.   The dust masses are integrated over all
the pixels that are detected at $>$3$\sigma$ in all observed bands.  The
total H gas masses given in the table 
footnote are integrated for the same pixels as the dust masses.  The
HI masses are directly from the HI measurements \citep{Stanimirovic00,
Muller03, Kim03} without any correction for opaque HI \citep{Dickey00,
Fukui14}.  The H$_2$
masses are computed from CO observations \citep{Mizuno01, Mizuno06,
Fukui08, Wong11} using $\mathrm{X}_\mathrm{CO} = 4.7 \times 10^{20}$
\citep{Hughes10} for the LMC and $\mathrm{X}_\mathrm{CO} = 6 \times
10^{21}$ \citep{Leroy07SMC} for the SMC.  The appropriate
$\mathrm{X}_\mathrm{CO}$ to use is a matter of debate, but the
expected range of this conversion factor is not large enough to
strongly impact the total gas masses \citep{Fukui10,
  Bolatto13}.  The ratios given only include
hydrogen, so are formally H gas-to-dust ratios, but for simplicity we
refer to them as gas-to-dust ratios.

The range of reasonable gas-to-dust ratios can be estimated three ways.
The first scales the range of observed gas-to-dust ratios in the Milky
Way by the LMC and SMC metallicities.  The second assumes the Milky
Way depletion factors and applies them to the measured LMC and SMC
abundances.  The third assumes all the metals available are in the
form of dust and this produces a minimum possible gas-to-dust ratio.
The MW depletions and gas-to-dust ratios vary
with environment and the global values in the Magellanic Clouds will
be some unknown mix of different ISM environments.  As a result, we
can only predict a possible range of gas-to-dust ratios.

The first method assumes that the relative amount of metals in the
LMC and SMC dust is the same as the MW, but scaled in proportion to
each galaxy's metallicity.  Thus, the expected gas-to-dust ratio 
will be 2X (LMC) and 5X (SMC) the MW gas-to-dust ratio.  The MW
gas-to-dust ratio varies from $\sim$250 for the very diffuse ISM ($F_*
= 0$) to $\sim$100 for the moderately dense ISM ($F_* = 1$) 
\citep{Jenkins09}.  For the LMC, we therefore expect a
gas-to-dust ratio between 200 to 500 while, for the SMC, we
expect a gas-to-dust ratio between 500 and 1250.  
The second method assumes the MW depletion patterns \citep{Jenkins09}
and the measured LMC and SMC abundances for each element \citep{Russell92}.  The
resulting expected LMC gas-to-dust ratios range between 150 to 
360 and the expected SMC gas-to-dust ratios range between 540 to
1300.  Combining the two different methods, the expected gas-to-dust
ratios are 150 to 500 and 500 to 1300 for the LMC and SMC,
respectively.  Finally, the minimum allowed gas-to-dust ratio can be
computed by assuming all the metals in the ISM in the form of dust.  Assuming
the measured LMC and SMC abundances, this gives minimum gas-to-dust
ratios of 105 and 300, respectively.  These expected gas-to-dust
ratios are given in Table~\ref{tab_dmasses}.

The gas-to-dust ratios for all three models are plotted in
Fig.~\ref{fig_gdrs} along with the allowed ranges for reasonable
depletions and maximum depletion.  From Table~\ref{tab_dmasses} and
this figure, it is clear that the TTMBB models give gas-to-dust ratios
that are lower than even possible assuming all the metals are present
in dust.  The TTMBB model gives low gas-to-dust ratios as it requires
large dust masses for the second cold component to be able to reproduce
the observed submm excess emission.
Thus, the TTMBB model is not a reasonable model for the dust
emission in the LMC or SMC.  The SMBB and BEMBB models give similar 
gas-to-dust ratios for both galaxies.  For the LMC, both models give
ratios that are well within the reasonable range of values.  For the
SMC, these two models both give values that are above the reasonable
values.  This is an indication that the depletions in the SMC are
lower than those the in MW or that the dust properties are
different (e.g. a smaller $\kappaeff$ value than that assumed in this
paper).

\subsection{Spatial Variations}

\begin{figure*}[tbp]
\epsscale{1.0}
\plotone{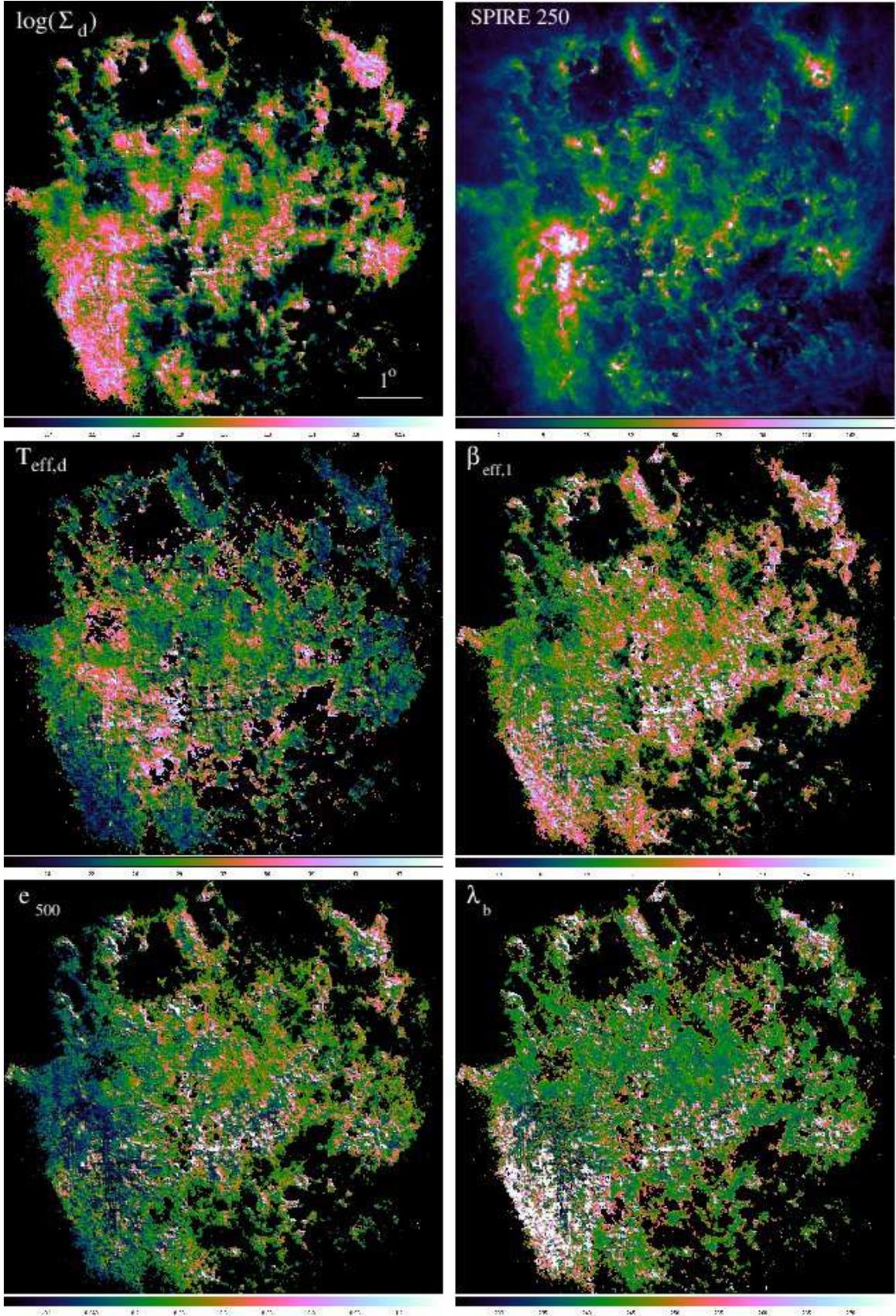}
\caption{The spatial distribution of $\log(\Sigma_d)$, $\dusttemp$,
  $\beta_\mathrm{eff,1}$, $e_\mathrm{500}$, and $\lambda_b$ for the
  BEMBB model are shown for the LMC using the 
  expectation value for each pixel.  In addition, the processed
  SPIRE~250~\micron\ image (\S\ref{sec_data}) is shown
  The images are shown using the
  cubehelix color mapping \citep{Green11}. 
  The left/right and up/down streaks seen are residual
  instrumental artifacts that are aligned along the PACS/SPIRE
  scan direction.
  \label{fig_lmc_images}}
\end{figure*}

\begin{figure*}[tbp]
\epsscale{1.0}
\plotone{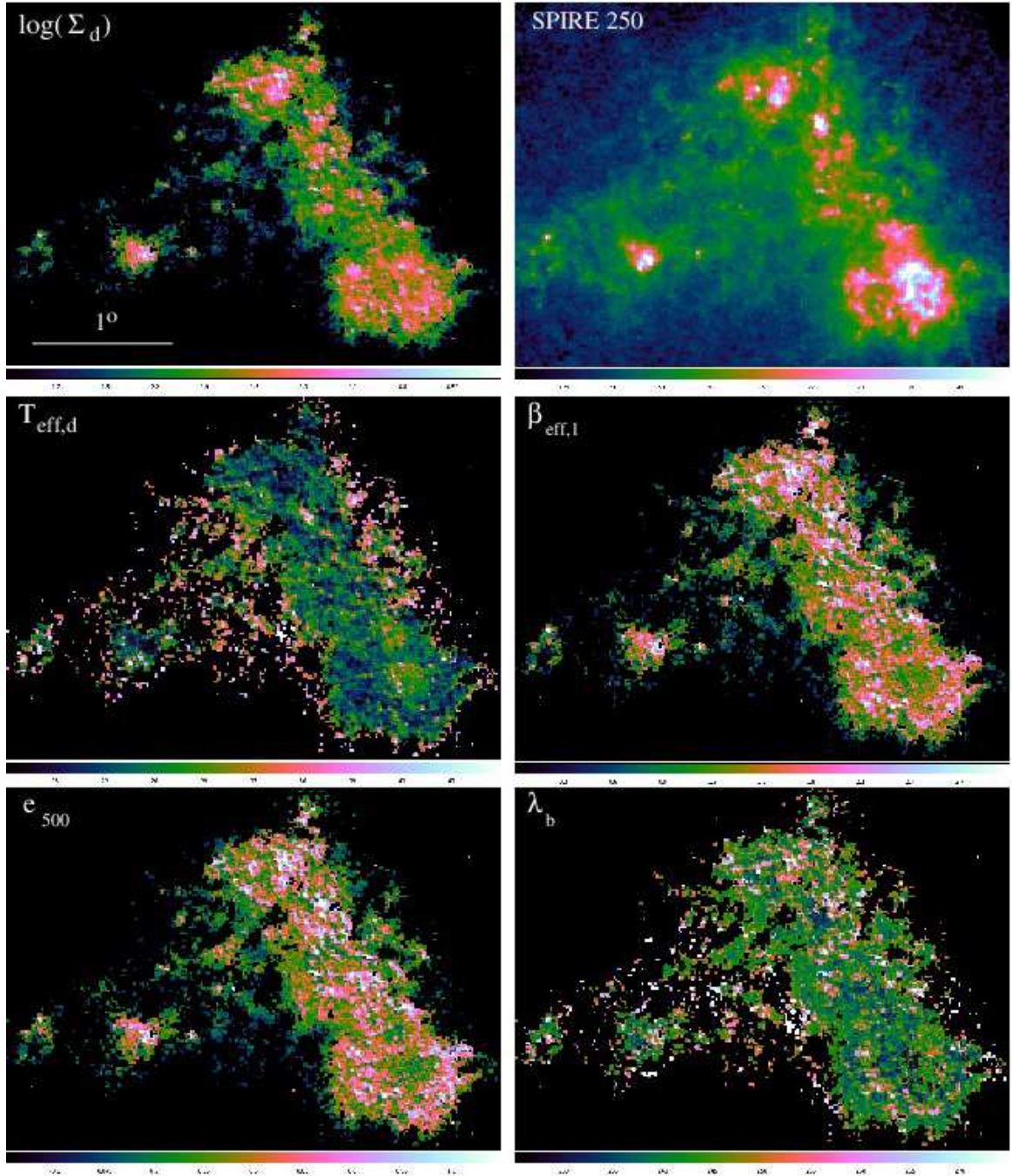}
\caption{The spatial distribution of $\Sigma_d$, $\dusttemp$,
  $\beta_\mathrm{eff,1}$, $e_\mathrm{500}$, and $\lambda_b$ for the
  BEMBB model are shown for the SMC using the `exp' value for each
  pixel.  In addition, the processed SPIRE~250~\micron\ image
  (\S\ref{sec_data}) is shown The images are shown using the cubehelix
  color mapping \citep{Green11}.
  \label{fig_smc_images}}
\end{figure*}

\begin{figure*}[tbp]
\epsscale{1.15}
\plotone{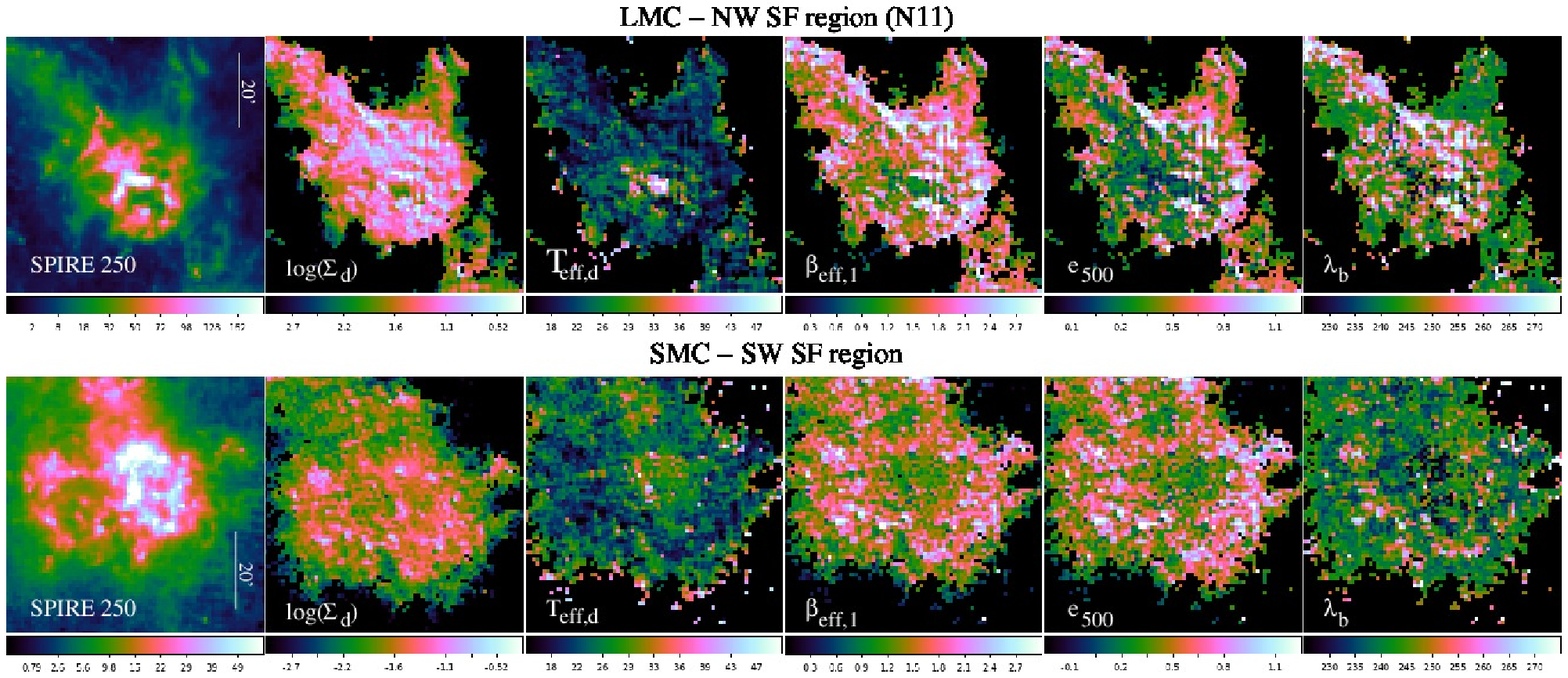}
\caption{The spatial distribution of $\Sigma_d$, $\dusttemp$,
  $\beta_\mathrm{eff,1}$, $e_\mathrm{500}$, and $\lambda_b$ for the
  BEMBB model are shown for one star forming region each in the LMC and
  SMC using the `exp' value for each pixel.  In addition, the
  processed SPIRE~250~\micron\ images (\S\ref{sec_data}) is shown The
  images are shown using the cubehelix color mapping \citep{Green11}.
  \label{fig_mc_images_closeups}}
\end{figure*}

The spatial variations across both galaxies in the different fit
parameters for the BEMBB 
model are shown in Figs.~\ref{fig_lmc_images} and
\ref{fig_smc_images}.  We only show the BEMBB results here as the
evidence in the previous subsections gives a fairly strong
indication that the BEMBB fits the data best
(\S\ref{sec_fit_residuals}) and provides a physically reasonable
gas-to-dust ratio (\S\ref{sec_gas_to_dust}).  The maps of dust surface
density ($\Sigma_d$) and temperature ($\dusttemp$) show qualitatively
similar behaviors to previous works \citep{Bot04, Leroy07,
  Bernard08}.  In detail, our maps differ mainly in showing finer
structure due to the higher spatial resolution {\it Herschel} observations.
One illustration of this effect is that the peak $\dusttemp$ in the
30~Dor region in our map is $\sim$60~K, significantly higher than
the $\sim$35~K found by \citet{Bernard08}.  

The higher spatial resolution of our maps does allow for detailed
investigations of individual star forming regions.  This is
illustrated by Fig.~\ref{fig_mc_images_closeups} where cutouts of the
BEMBB fit parameter maps for a star forming
region in each galaxy are shown.  The morphology of these two star
forming regions is similar.  The SPIRE 250~\micron\ emission is
strongly peaked in the region centers in contrast to the dust surface
density which is more constant across the regions.  This difference
is caused by the center of these regions having high $\dusttemp$
values.  The $\betaeff$ and $e_\mathrm{500}$ maps of both regions have
very similar morphologies, visually illustrating that these two fit
parameters are strongly correlated.  Finally, the $\lambda_b$ images
show coherent structures with fairly small variations overall. 
The submm excess as parametrized by $e_\mathrm{500}$ is near zero in
the center of the two star forming regions and rises rapidly to values
around one near the edges.  This behavior is
intriguing, but the strong correlations of $e_\mathrm{500}$ with
$\betaeff$ indicate that 
more work is needed to determine if this is real or due to noise
induced correlations.  

The overall properties of the global submm excess between the LMC and
SMC show trends that are consistent with previous work.  The average
LMC and SMC $e_\mathrm{500}$ values are 0.27 and 0.43 when the average
is done using the `realize' method and each pixel has equal weight.
This can be visually seen in the $e_\mathrm{500}$ images in
Figs.~\ref{fig_lmc_images} and \ref{fig_smc_images} where the SMC
shows a higher filling factor of high $e_\mathrm{500}$ values than the
LMC.  This trend of the lower metallicity SMC having a higher submm
excess than the LMC is expected given the results from global studies
of the submm excess \cite{RemyRuyer13}.  A fairer comparison of the
absolute value of $e_\mathrm{500}$ with global SED fits is the dust
surface density weighted averages that are 0.11 and 0.26 for the LMC
and SMC, respectively.  Finally, the average values of $\lambda_b$ are
$\sim$240 for both types of averages and both galaxies.  This
wavelength is similar to that found by \citet{Li01} from fitting the
DIRBE MW diffuse spectrum.

\begin{figure*}[tbp]
\epsscale{1.3}
\plotone{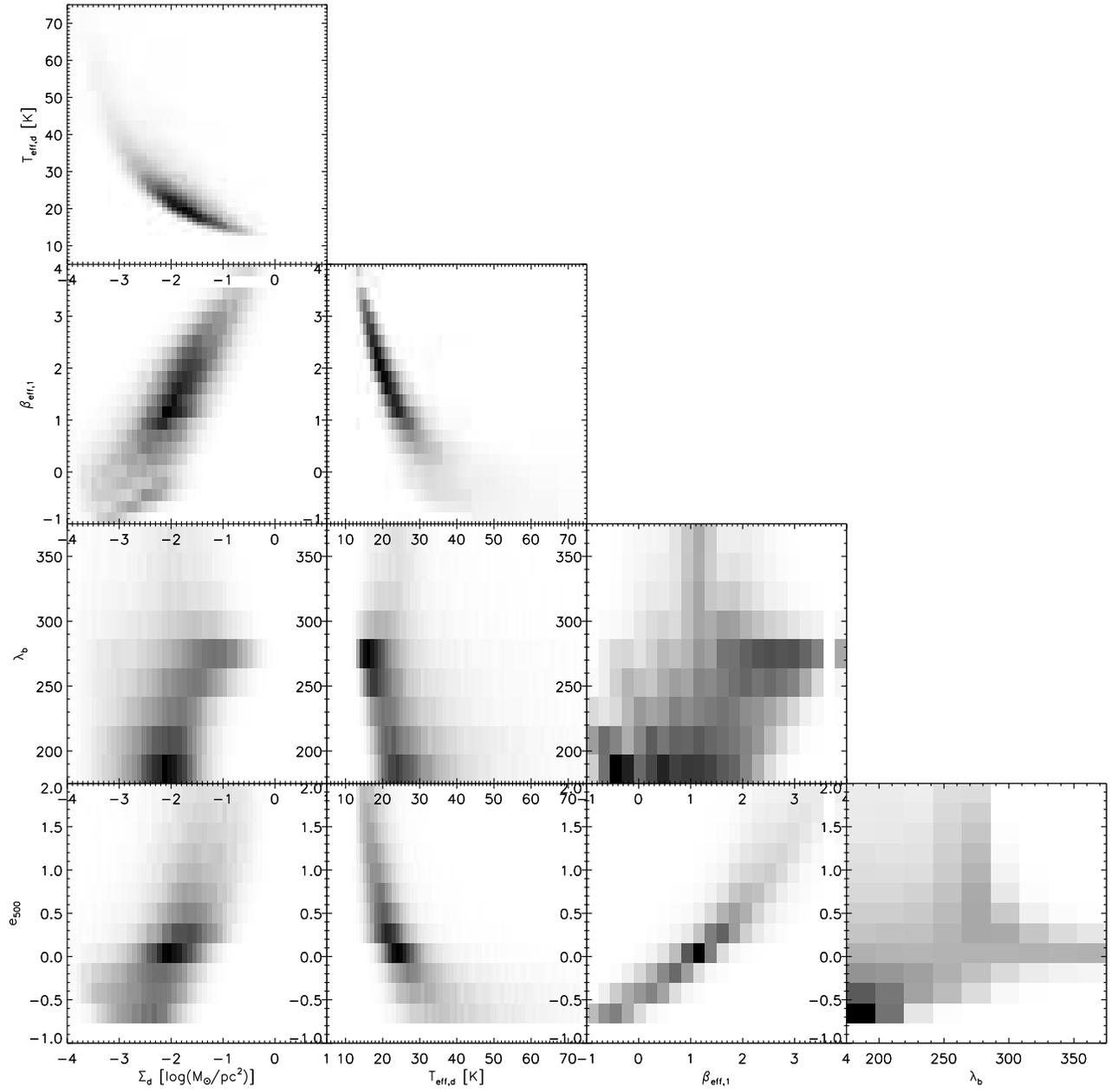}
\caption{The correlations for the LMC between all the five fit
  parameters for the BEMBB model are plotted.  The plots are density
  plots where each point that contributes to the density is a single
  realization of the full likelihood function for a single pixel.
  \label{fig_lmc_param_trends}}
\end{figure*}

To investigate the variations in fit parameters more quantitatively,
we plot all the correlations between the different fit parameters for
the LMC in Fig.~\ref{fig_lmc_param_trends}.  The plots for the SMC are
very similar and are not shown.  These plots show
the density of points where each point represents a single pixel.  The
values used for each pixel use the `realize' method where the
likelihood functions are randomly sampled once for each pixel.  This
means that these density plots statistically sample the full
information for the fit from each pixel.  Repeating the `realize'
method process with a different random sampling for each pixel
produces plots that are very similar.  This indicates that these plots
fully capture the correlations between fit parameters with a single
sampling of each pixel's likelihood function due to the large number
of pixels.  Plots created using the `max' and `exp' methods are
significantly different as they do not fully include the information
on the uncertainties in the fits to each pixel.  As an example of the
difference between the different ``best fit'' methods, a flat
likelihood function would show a single value for `max' and `exp',
while the `realize' method would have a value that was randomly
distributed over the entire parameter range.

These plots show that many of the parameters are correlated with each
other, sometimes quite strongly.  The strongest correlations are seen between
$\log(\Sigma_d)$ and $\dusttemp$, $\log(\Sigma_d)$ and $\beta_\mathrm{eff,1}$,
$\dusttemp$ and $\beta_\mathrm{eff,1}$, and $\beta_\mathrm{eff,1}$ and
$e_\mathrm{500}$.  The origin of these correlations can
be either real or a result of interactions between noise in the
measurements and model fit parameters.  The correlation between
$\log(\Sigma_d)$ and $\dusttemp$ is real in that it reflects the
detection thresholds of the HERITAGE data.  Hotter dust can be
detected a lower dust surface densities than cooler dust due to the
$\dusttemp^4$ behavior of blackbodies.  The anti-correlation between
$\dusttemp$ and $\betaeff$ is one of the correlations that has been
studied extensively to learn if it is due to noise or real variations
in the dust properties \citep{Dupac03, Shetty09a, Shetty09b,
  Galliano11, Juvela12, Kelly12, Ysard12, Veneziani13}.  Laboratory
data on dust analogs do show a shallow anti-correlation between
$\dusttemp$ and $\betaeff$ \citep{Coupeaud11}, but noise in
measurements also produces a similar or larger anti-correlation.
\citet{Kelly12} have proposed to use a hierarchical Bayesian model to
solve for the true $\dusttemp$--$\betaeff$ correlation, where the
hierarchical model assumes a single $\dusttemp$ and $\betaeff$ with
some distribution around these values.  In fitting an entire galaxy,
such an assumption is not justified as, for example, there are regions
near star formation that will be significantly hotter than regions
further away.  In addition, \citet{Juvela13} find there are biases in
all the currently proposed methods for determining the true
$\dusttemp$--$\betaeff$ relation.  Thus, we choose to graphically
display the correlations using the `realize' method and not explicitly
fit for the correlation.  In future work, we plan to incorporate
additional observations of the ISM and physical models for the
correlations between different ISM parameters (e.g. dust and gas
surface densities).

Fig.~\ref{fig_lmc_param_trends} shows the 
correlations between the submm excess $e_\mathrm{500}$ and other dust
properties.  The value of $e_\mathrm{500}$ is positively correlated
with $\Sigma_d$ and $\beta_\mathrm{eff,1}$ and negatively correlated
with $\dusttemp$.  This may be real or it may be due to the
$\dusttemp$ versus $\beta_\mathrm{eff,1}$ anti-correlation that is
also very clearly seen.  The positive correlation between
$e_\mathrm{500}$ and $\Sigma_d$ is the opposite of what was found by
\citet{Galliano11} for a pathfinder study using a portion of the
HERITAGE data on the LMC and \citet{Paradis12} for the MW.  The
difference between these works and our work may be due to changes in
the PACS and SPIRE calibration, different fitting methods, and/or
different dust emission models.  Future work will investigate these
differences by using the same data, same fitting code, and expanding
the dust emission model to include more sophisticated dust emission models.

\section{Conclusions and Future}

We find that the Magellanic Clouds show a submm excess in the {\it
  Herschel} HERITAGE observations with a spatial resolution of
$\sim$10~pc.  This submm excess seen in the Magellanic Clouds is more
likely to be due to variations in the dust emissivity wavelength
dependence than a second population of colder dust.  This is based on
the BEMBB model providing the best fit to the HERITAGE data and
producing realistic gas-to-dust ratio values.  The average submm
excesses seen at 500~\micron\ at $\sim$10~pc resolution are 27\% and
43\% for the LMC and SMC, respectively.  There are trends of the submm
excess and environment (probed by $\Sigma_d$ and $\dusttemp$), but the
true nature of these trends will be investigated in future work
incorporating more data and more physical models of the ISM.

The total dust masses integrated over the pixels detected at $3\sigma$
in all five PACS/SPIRE bands using our favored model (BEMBB) are $(7.3
\pm 1.7) \times 10^5$ and $(8.3 \pm 2.1) \times
10^4$~$\mathrm{M}_\sun$ for the LMC and SMC, respectively.  These dust
masses are significantly lower (factors of 4--5) than would be
expected from previous dust masses measurements \citep{Leroy07,
Bot10}.  The lower dust masses we derive have important implications
for the study of the lifecycle of dust in the Magellanic Clouds as the
relative contributions between Asymptotic Giant Branch (AGB),
supernove, and the ISM for the formation of dust change significantly
\citep{Matsuura09, Boyer12, Matsurra13, Zhukovska13}.

Future work will focus on adding more physics to the fitting for dust
properties.  One rich area for future work will be to include
constraints from other observations of the ISM in the Magellanic
Clouds.  An initial foray into this area is the focus of Roman-Duval
et al. (this issue) who use the dust surface densities from this paper
to investigate the dependence of the gas-to-dust ratio on environment.
For the dust modeling in particular, future work will include more
sophisticated dust grain models \citep[e.g.][]{Weingartner01,
Compiegne11, Galliano11} and shorter wavelength infrared observations (e.g.,
Spitzer IRAC/MIPS data) to better constrain the possible grain
compositions.

\acknowledgements
We greatly benefited from conversations with Morgan Fouesneau, David
Hogg, Derck Massa, and Daniel Weisz on the always interesting topic of
fitting data with models.  We acknowledge financial support from the
NASA {\it Herschel} Science Center, JPL contracts \# 1381522 \&
1381650.  M.R. acknowledges partial support from CONICYT project BASAL
PFB-6.

\end{document}